\documentstyle[aps,prb,eqsecnum,preprint,tighten,amssymb,epsf,floats]{revtex}

\makeatletter
\def\p@subsection{\thesection.}
\makeatother
\begin{document}

\draft
\title{Critical Dynamics of Gelation}

\author{Kurt Broderix,\thanks{Deceased (12 May 2000)}
        Henning L\"owe, Peter M\"uller, and Annette Zippelius}

\address{Institut f\"ur Theoretische Physik,
         Georg-August-Universit\"at, D--37073 G\"ottingen, Germany}
\date{Version of 4 September 2000}
\maketitle

\begin{abstract}
  Shear relaxation and dynamic density fluctuations are studied within
  a Rouse model, generalized to include the effects of permanent
  random crosslinks. We derive an exact correspondence
  between the static shear 
  viscosity and the resistance of a random resistor network.
  This relation allows us to compute the static shear
  viscosity exactly for uncorrelated crosslinks. For more general
  percolation models, which are amenable to a scaling description, it
  yields the scaling
  relation $ k=\phi-\beta$ for the critical exponent of the shear
  viscosity. Here $\beta$ is the thermal exponent for the gel fraction
  and $\phi$ is the crossover exponent of the resistor network. 
  The results on the shear viscosity are also used in deriving  upper and lower
  bounds on the incoherent scattering function in the long-time limit,
  thereby corroborating previous results. 
\end{abstract}


\narrowtext

\section{Introduction}\label{Sec1}

\begin{figure}[t]
  \center \leavevmode
  \epsffile{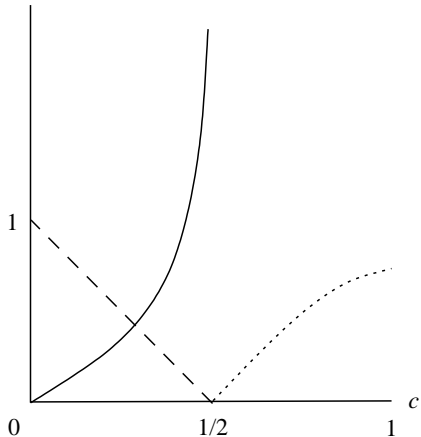} 
  \medskip
  \caption{\label{fig1} Sketch of some critical quantities for 
    gelation as a function of the crosslink concentration $c$
    for a mean-field distribution of crosslinks. The gelation
    transition occurs at $c_{{\rm crit}}=1/2$. The plot displays the
    averaged static shear viscosity (solid line) according to Eq.\
    (\protect\ref{Eq3.24}), the effective diffusion constant (dashed
    line) according to Eq.\ (\protect\ref{deeff2}) and the gel
    fraction (dotted line) according to
    Ref.~\protect\onlinecite{BrGoZi97}.}   
\end{figure}

The most prominent critical phenomena at the gelation or vulcanization
transition are of dynamic origin, see e.g.\ Ref.\ \onlinecite{MaAd91}
for a recent 
review. Relaxation times in the sol phase diverge as the 
gelation transition is approached,\cite{MaWi88-AdMa90-MaWiOd91}
giving rise to critical behaviour of the transport coefficients: The
effective diffusion constant goes to zero and the static shear
viscosity diverges,
\cite{AdDeDuHiMu81-AdDeDu85,MaWi88-AdMa90-MaWiOd91} indicating
structural arrest of a macroscopic fraction of the monomers; see
Figure~\ref{fig1} below for a graphical illustration of the
situation. There are 
strong precursors of the gelation transition in the fluid-like sol
phase: The decay of correlations is not exponential, but follows a
Kohlrausch law.\cite{MaWi88-AdMa90-MaWiOd91} This indicates a spectrum
of relaxation times extending to arbitrary large values.

All these phenomena are reminiscent of the glass transition so that the
gelation transition can be regarded as a paragon for the latter.  There are
however important differences which are best explained in the context of
vulcanization. In that case the distinction between thermal and quenched
degrees of freedom is simple and unambiguous: Chemical crosslinks are quenched
and the monomers equilibrate in a fixed crosslink configuration. In contrast to
this scenario, the quench in the structural glass by lowering the temperature
does not allow for such a simple classification. Even though it is generally
believed that while decreasing the temperature at a finite rate $\omega$ some
degrees of freedom are quenched, namely those whose relaxation times $\tau$ are
larger than the inverse quench rate, $\tau \omega > 1$.  Nevertheless the
identification of these quenched degrees of freedom is neither simple nor
unique, and some properties of structural glasses do in fact depend on the rate
of quench.\cite{VoKo96} The so-called physical gelation seems a
better candidate to model glassy 
behaviour. In that case the temperature can be easily adjusted so that the
binding energy for two monomers participating in a crosslink is comparable to
the thermal energy. Crosslinks form and break up so that both monomer
positions and crosslinks equilibrate.  By lowering the temperature crosslinks
are quenched and give rise to structural arrest at sufficiently low
temperatures.

In this paper we will concentrate on vulcanization and chemical gelation, 
both of which are commonly interpreted as a percolation 
transition.\cite{StCoAd82} The transition from a fluid phase to an amorphous
solid phase happens when a macroscopic cluster of crosslinked polymers
has been formed.  However, percolation theory can only account for the
geometric connectivity of the macromolecules and neither thermal
fluctuations nor dynamic phenomena are comprised in the percolation
picture. Instead, one should start from a dynamic model, as is done
here. The simplest model, Rouse dynamics, ignores all 
interactions except for those which ensure the connectivity of the
clusters. De Gennes\cite{Gen78} was the first to estimate the static
shear viscosity near the gelation transition. Relating the viscosity
$\eta(n)$ of a cluster of mass $n$ to the longest relaxation time of
the cluster, he argues that $\eta(n)\sim R^2$ scales like the squared
linear dimension of the cluster. The radius of gyration is determined
by the mass of the cluster according to $R\sim n^{1/d_f}$, where
$d_f=d-\beta / \nu$ is the Hausdorff-Besicovitch dimension of the
fractal and related to the exponent for the correlation length $\nu$
and the gel fraction $\beta$.  Given a scaling ansatz \cite{StAh94}
for the number of $n$-clusters per polymer, $\tau_n\sim n^{-\tau}
\exp\!\left\{-n/n^{*}\right\}$ with $n^{*}\sim |p-p_c|^{-1/\sigma}$,
one can average 
over all cluster sizes to obtain $\eta=\sum_n \eta (n) n \tau_n
\sim|p-p_c|^{-k}$ with $k=2 \nu - \beta$.  This scaling relation was
rederived in many other ways. It is incorrect because it ignores the
internal structure of percolation clusters which is not only
determined by the exponents $\beta$ and $\nu$. The latter rule the
behaviour of clusters on large spatial scales, whereas the intrinsic
self-similar connectivity is characterized by the spectral dimension
$d_s$. In contrast to a linear chain with $d_s=1$, a percolation
cluster contains a hierarchy of branches and possibly loops, giving
rise to a value of $d_s>1$.  
Apparently, Cates \cite{Ca85} was the
first to notice 
that the relaxation spectrum acquires a Lifshits tail which modifies
the above estimate for the viscosity $\eta(n)$ of a single percolation
cluster.

In a different approach the viscosity at the sol-gel transition has
been related to the conductivity of a random resistor network
\cite{Ge79} which is made up of a fraction $p$ of superconducting and
a fraction $(1-p)$ of normal conducting bonds. Approaching the
percolation transition from below, the conductance $\sigma \sim
|p-p_c|^{-s}$ is expected to diverge with an exponent $s$, which was
predicted to be the same as $k$.  Before it was realized that
percolation clusters are multifractals, the two scaling arguments were
considered to give identical results, namely $k=s= 2\nu -\beta$. If
multifractality is treated properly, the two scaling arguments give
different results.  This is most easily seen in two dimensions, where
duality \cite{Str77} implies $s=\phi$.  Here, $\phi$ is the so-called
crossover exponent, which cannot be expressed through $\beta$ and
$\nu$. It was computed by Harris and Lubensky 
\cite{HaLu87} in
the context of random resistor networks where a fraction $1-p$ of normal
conducting bonds has been removed. The crossover exponent $\phi$
characterizes the average resistance $\Omega(r)\sim r^{\phi/\nu}$
between two connected nodes which are a distance $r$ apart.

In this paper we concentrate on the fluid-like sol phase of a gelling
polymer melt. We show that for Rouse dynamics the exponent of the
static shear viscosity $k$ is given exactly by
\begin{equation}\label{Eq1.1}
  k=\phi-\beta
\end{equation}
in disagreement with {\it both} of the above scaling arguments. The derivation
of the exponent $k$ requires two steps. First, the viscosity is expressed in
terms of the random connectivity matrix, which characterizes the
connections in the polymer
network. In a second step we prove that bonds between monomers can be
identified with electrical
resistors and use results from the theory of random graphs to relate matrix
elements of the connectivity matrix to the resistance between two
nodes $i$ and $j$ of the corresponding random resistor network. In
fact, we consider the derivation of such a correspondence as one of
the main results of this paper.

For a mean-field distribution of crosslinks
we compute the static shear viscosity
exactly and find a logarithmic divergence, in agreement with the
general scaling 
relation (\ref{Eq1.1}) for mean-field exponents or space dimensionality $d\geq
6$. For a crosslink distribution which corresponds to
finite-dimensional percolation, 
we know the $\epsilon$-expansion of $k=\epsilon/6+11\epsilon^2/1764+{\cal
  O}(\epsilon^3)$ or use high-precision simulations in $d=3$ to obtain
$k|_{d=3} \approx 0.71$.

The established correspondence between the random polymer network and
the random resistor network turns out to be useful also for the
computation of the intermediate incoherent scattering function
$S_t(q)$. The long-time asymptotics of $S_t(q)$ was discussed in Ref.\
\onlinecite{BrGoZi97} for a mean-field distribution of crosslinks. In
this paper, we 
derive upper and lower bounds on $S_t(q)$ which put our previous
results on a firmer basis. We furthermore use a scaling analysis to
extract the long-time asymptotics of $S_t(q)$ for
a crosslink distribution which corresponds to finite-dimensional
percolation. 

The paper is organized as follows: In the following Section~\ref{Sec2} we
introduce the basic dynamic model of a monodisperse sol of phantom monomer
chains, which are permanently crosslinked by Hookean springs chosen at random.
\cite{SoVi95} We present the formal solution of the dynamic model and
introduce the two observables, which we want to discuss: the stress relaxation
function and the incoherent scattering function. In Section~\ref{Sec3} 
we discuss the static shear viscosity. First, the relation between our
dynamic model and random resistor networks is established. Second, we
relate the critical exponent of the viscosity to the exponents arising 
in a scaling description of the crosslink distribution,
as exists for finite-dimensional percolation. Third, we show how to
compute the static shear viscosity exactly for the case where the
crosslinks are distributed according to mean-field percolation. 
This result is then rederived using replicas in the last subsection of
Section~\ref{Sec3}.
The intermediate incoherent scattering function is discussed in 
Section~\ref{Sec4}. We first
derive exact upper and lower bounds, which are then used to extract
the long-time asymptotics for a mean-field distribution of crosslinks
and corroborate previous results. 
\cite{BrGoZi97} Some details of this calculation are delegated to the
Appendix. In the last part of Section~\ref{Sec4} we compute critical
exponents associated with the 
long-time asymptotics of the 
scattering function for a crosslink distribution which admits a
scaling description---as is the case for finite-dimensional 
percolation. We conclude with
some remarks and perspectives for 
future directions in Section~\ref{Sec5}. A brief account of our
results was given in Ref.\ \onlinecite{BrLoMuZi99}. 

%
%

\section{Dynamic Model and its solution}\label{Sec2}

\subsection{Dynamic Model}\label{Sec2.1}

We consider a system of $N$ identical, mono-disperse (macro-)molecular units,
each consisting of $L$ monomers. These units may be monomers, dimers, linear
chains, or arbitrary branched structures, which may also contain
loops. In the following these molecular units are 
called polymers, regardless of their topology or their degree of
polymerization $L$. Monomer $s$ 
on polymer $i$ is characterized by its time-dependent position vector ${\bf
  R}_t(i,s)$ ($i=1,\ldots,N$ and $s=1,\ldots,L$) in $d$-dimensional space.  We
suppose that the polymers are subjected to a space- and time-dependent external
velocity field $v^{\alpha}_t({\bf r})$. Here, Greek indices indicate Cartesian
co-ordinates $\alpha=x,y,z,\dots$, and we will always consider a flow field in
the $r^x$-direction, increasing linearly with $r^y$, {\it i.e.}
\begin{equation}\label{Eq2.1}
  v^{\alpha}_t({\bf r}) :=
  \delta_{\alpha,x} \kappa_t r^y,
\end{equation}
with a time-dependent shear rate $\kappa_t$.  As usual, $\delta_{\alpha,\beta}$
denotes the Kronecker symbol, {\it i.e.}\ $\delta_{\alpha,\beta}=1$ for
$\alpha=\beta$ and zero otherwise.

We employ the simplest, purely relaxational dynamics, see e.g.\
Chap.~4 in Ref.\ \onlinecite{DoEd85} or Chap.~15.1 in Ref.\
\onlinecite{BiCu87} 
\begin{equation}\label{Eq2.2}
  \partial_t R^{\alpha}_t(i,s) = 
  - \frac{1}{\zeta} \: \frac{\partial H}{\partial R^{\alpha}_t(i,s)} 
  + v^{\alpha}_t\bigl({\bf R}_t(i,s)\bigr) + \xi^{\alpha}_t(i,s).
\end{equation}
In the absence of the thermal noise $\bbox{ \xi}$ the monomers relax to
the state $\partial H/\partial{\bf R}={\bf 0}$ such that their velocity is
equal to the externally imposed velocity field. The noise $\bbox{ \xi}$
has zero mean and covariance
$\overline{\xi^{\alpha}_t(i,s)\,\xi^{\beta}_{t'}(i',s')} = {2}{\zeta}^{-1}\,
\delta_{\alpha,\beta}\, \delta_{i,i'}\, \delta_{s,s'}\, \delta(t-t')$, where
$\delta(t)$ is the Dirac $\delta$-function. Here, the overbar indicates the
average over the realizations of the Gaussian noise $\bbox{\xi}$.  The
relaxation constant is denoted by $\zeta$, and we use energy units such that
$k_B T=1$. In (\ref{Eq2.2}) the Hamiltonian $H:=H_W+U$ consists
of two terms. The first one guarantees the connectivity of each polymer, whose
internal structure is characterized by a connectivity matrix
$\Gamma_{{\rm poly}}(s;s')$,
\begin{equation}\label{Eq2.3}
  H_W := \frac{d}{2l^2}\sum_{i=1}^{N}\sum_{s,s'=1}^{L} 
  \Gamma_{{\rm poly}}(s;s')\, {\bf R}(i,s) \cdot {\bf R}(i,s') \,.
\end{equation}
We have chosen a harmonic potential to constrain the relative distance between
two monomers on the same polymer with the typical distance given by the
persistence length $l>0$. As an example for $\Gamma_{{\rm poly}}$ we mention
the special case of linear chains. Their connectivity matrix is explicitly
given by
\begin{equation}\label{Eq2.4}
  \Gamma_{{\rm poly}}(s;s') 
  = \sum_{r=1}^{L-1} (\delta _{s,r} - \delta _{s,r+1})\,
    (\delta _{s',r} - \delta _{s',r+1})\,.
\end{equation}
The second part of the Hamiltonian models $M$ permanently formed crosslinks
between randomly chosen pairs of monomers. The configuration of crosslinks will
be denoted by ${\cal G}=\{i_e,s_e;i'_e,s'_e\}_{e=1}^M$, {\it i.e.}\ the list of
the $M$ crosslinked pairs of monomers.  The relative distances between any two
monomers, participating in a crosslink, are constrained by a harmonic potential
\begin{equation}\label{Eq2.5}
  U := \frac{d}{2a^2}\:\sum_{e=1}^M 
  \bigl( {\bf R}(i_e,s_e)-{\bf R}(i'_e,s'_e) \bigr)^2,
\end{equation}
whose strength is controlled by the parameter $a>0$. For $a\to 0$ hard
crosslinks can be recovered. \cite{SoVi95} Since the Hamiltonian is quadratic
in the monomer positions, it can be expressed in terms of a $NL\times NL$
connectivity matrix $\Gamma$ according to
\begin{equation}\label{Eq2.6}
  H=: \frac{d}{2a^2}\:\sum_{i,i'=1}^{N}\sum_{s,s'=1}^{L} \Gamma(i,s;i',s')\,
  {\bf R}(i,s)\cdot {\bf R}(i',s')  \,.
\end{equation}
The connectivity matrix $\Gamma$ has a deterministic part
$\Gamma_{{\rm poly}}$, which reflects the internal structure of the polymers,
and a random part representing the crosslinks
\widetext
\begin{equation}
  \label{Eq2.7}
  \Gamma(i,s;i',s') 
  =  \frac{a^2}{l^2}\:\delta_{i,i'} \Gamma_{{\rm poly}}(s,s')  
  + \: \sum_{e=1}^{M} 
  (\delta_{i,i_{e}}\delta_{s,s_{e}} - \delta_{i,i_{e}'}\delta_{s,s_{e}'})
  (\delta_{i',i_{e}}\delta_{s',s_{e}} - 
  \delta_{i',i_{e}'}\delta_{s',s_{e}'})\,. 
\end{equation}
\narrowtext  
In order to determine the model completely, it only remains to specify the
probability distribution of the crosslink configurations. We shall discuss
two different types of probability distributions $P({\cal G})$:\\
$i)$ crosslinks are chosen independently with equal probability for every pair
of monomers, corresponding to a mean-field distribution,
\cite{ErRe60,Ste77}\\
$ii)$ a distribution of crosslinks, which generates clusters amenable to the
scaling description of finite-dimensional percolation.\\
The precise characterization of these distributions is given below.
\subsection{Observables}\label{Sec2.2}
We shall focus on two observables. First, we aim at the computation of the
intrinsic shear stress $\sigma^{\alpha,\beta}_t$ as a function of the shear
rate $\kappa_t$.  Following Chap.~3 in Ref.\ \onlinecite{DoEd85} or 
Chap.\ 16.3 in Ref.\ \onlinecite{BiCu87}, we express the shear stress
in terms of the force per unit area, exerted by the polymers
\begin{equation}\label{Eq2.8}
  \sigma^{\alpha,\beta}_t = \lim_{t_{0}\to-\infty}
  -\frac{\rho_{0}}{N}\sum_{i=1}^{N}\sum_{s=1}^L 
  \overline{F^{\alpha}_t(i,s) R^{\beta}_t(i,s)}.
\end{equation}
Here, ${\bf R}_{t}(i,s)$ is the solution of the equation of motion
(\ref{Eq2.2}) with some initial condition ${\bf R}_{t_{0}}(i,s)$ at time
$t_{0}$ in the distant past.  Moreover, $\rho_{0}$ stands for the polymer
concentration and $F^{\alpha}_t(i,s):=-\partial H/\partial R^{\alpha}_t(i,s)$
is the force acting on monomer $(i,s)$ at time $t$.  For the simple shear flow
(\ref{Eq2.1}), a linear response relation
\begin{equation}\label{Eq2.9}
  \sigma^{x,y}_t=
  \int_{-\infty}^t d \tau\:G_{t-\tau}\,\kappa_{\tau}
\end{equation}
with response function $G_t$ will be shown to be valid for arbitrary strengths
of the shear rate $\kappa_t$. For a time-independent shear rate
$\kappa$ the shear stress is also independent of time and the 
intrinsic shear viscosity $\eta$ is then related to the shear stress via
\begin{equation}\label{Eq2.10}
  \eta({\cal G}) :=\sigma^{x,y}/(\kappa\rho_{0})=
  \rho_{0}^{-1}\int_0^{\infty}\! d  t \: G_t.
\end{equation}
Note that we have made the dependence of $\eta$ on the realization ${\cal G}$
of the crosslinks explicit.

Second, in the absence of the shear flow, $\kappa=0$, we compute the
intermediate incoherent scattering function
\widetext
\begin{equation}\label{Eq2.11}
  S_{t}(q|{\cal G}):=\lim_{t_{0}\to -\infty}
  \overline{
    \frac{1}{NL}
    \sum_{i=1}^N
    \sum_{s=1}^L
    \exp\!\left\{{i}\,{\bf q}\cdot\bigl({\bf R}_{t}(i,s)-{\bf
        R}_{0}(i,s)\bigl)\right\}}
\end{equation}
\narrowtext
with special emphasis on its long-time behaviour.
Due to isotropy the scattering function depends only on the modulus
$q:=|{\bf q}|$ of the wave vector. 

We expect that observables, like the viscosity and the incoherent scattering
function, are self averaging in the macroscopic limit and compute the averages
\begin{equation}
  \left\langle \eta\right\rangle := \sum_{{\cal G}}P({\cal G})\,\eta({\cal G})
\end{equation}
and, accordingly, $\left\langle S_t(q)\right\rangle$ over all
crosslink realizations.

\subsection{Formal Solution}\label{Sec2.3}
The equation of motion (\ref{Eq2.2}) is linear for the spatially homogeneous
shear gradient (\ref{Eq2.1}) and hence can be solved exactly.  For notational
convenience we introduce $NL$-dimensional vectors such as
$R_t^{\alpha}:=\bigl(R_t^{\alpha}(1,1),\ldots,R_t^{\alpha}(N,L)\bigl)$, whose
$NL$ components are the respective spatial components of the position vectors
${\bf R}_t(i,s)$.  The force vectors $F_t^{\alpha} = - (d/a^{2})\Gamma
R_{t}^{\alpha}$ and the noise vectors 
$\xi_t^{\alpha}$, $\alpha=x,y,z,\dots$, are defined in an analogous manner.
Furthermore we use the abbreviation $K^{\alpha,\beta}_t=\delta_{\alpha,
  x}\delta_{\beta, y}\kappa_t$ for the spatial components of the velocity
gradient tensor, which is according to (\ref{Eq2.1}) spatially homogeneous.

The expression (\ref{Eq2.8}) for the stress tensor can be rewritten as
\begin{eqnarray}\label{Eq2.12}
  \sigma^{\alpha,\beta}_t &=& \frac{\rho_0\, d}{Na^{2}}\: \lim_{t_{0}\to -\infty} 
  {\rm Tr}\bigl(\Gamma C^{\alpha,\beta}_t\bigl) \nonumber\\
  &=& 
  \frac{\rho_0\, d}{Na^{2}}\: \lim_{t_{0}\to -\infty}
  \sum_{i=1}^{N}\sum_{s=1}^{L} (\Gamma C_{t}^{\alpha, \beta})(i,s;i,s)
  \,,
\end{eqnarray}
where $C^{\alpha,\beta}_t :=
\overline{R^{\alpha}_t\bigl(R^{\beta}_t\bigr)^{+}}$ is the matrix of
second moments of ${\bf R}_t$ and the superscript~$^{\!\!+}$ denotes
transposition.  The matrix $C^{\alpha,\beta}_t$ may be calculated from the
equation of motion (\ref{Eq2.2}) as follows: Introducing
$U_t(i,s;i',s'):=\exp\!\left\{-d\,t / (\zeta
  a^2)\:\Gamma\right\}(i,s;i',s')$, it is readily verified that
\begin{equation}\label{Eq2.13}
  R^{\alpha}_t =
   U_{t-t_0} \sum_{\beta} T^{\alpha,\beta}_{t,t_0} R^{\beta}_{t_0}
  +\int_{t_0}^t d  t'\: 
   U_{t-t'} \sum_{\beta} T^{\alpha,\beta}_{t,t'} \xi^{\beta}_{t'}
\end{equation}
is the unique solution of (\ref{Eq2.2}), with initial condition ${\bf
  R}_{t_0}$, provided that $T^{\alpha,\beta}_{t,t'}$ is the solution of the
differential equation $\partial_t\: T^{\alpha,\beta}_{t,t'} =
\sum_{\gamma}K^{\alpha,\gamma}_t\:T^{\gamma,\beta}_{t,t'}$ with
initial condition
$T^{\alpha,\beta}_{t,t}=\delta_{\alpha,\beta}$.  Since $\bbox{ \xi}_t$
is Gaussian white noise with zero mean, we obtain
\begin{eqnarray}\label{Eq2.14}
  C^{\alpha,\beta}_t &=& 
    U_{2(t-t_0)} \sum_{\gamma,\gamma'} T^{\alpha,\gamma}_{t,t_0} 
    C^{\gamma,\gamma'}_{t_0} T^{\beta,\gamma'}_{t,t_0} \nonumber\\
  &&+ \frac{2}{\zeta}\int_{t_0}^{t}  d  t' \: U_{2(t-t')} \:
    \sum_{\gamma} T^{\alpha,\gamma}_{t,t'} \, T^{\beta,\gamma}_{t,t'}.
\end{eqnarray}
Multiplying Eq.\ (\ref{Eq2.14}) by $\Gamma$ and fixing the initial condition at
time $t_0\to -\infty$ we end up with
\begin{equation}\label{Eq2.15}
  \lim_{t_{0}\to -\infty} \Gamma \, C^{\alpha,\beta}_t =
  \frac{2}{\zeta}
  \int_{-\infty}^t d  t'\: \Gamma\,U_{2(t-t')}
  \sum_{\gamma} T^{\alpha,\gamma}_{t,t'} \, T^{\beta,\gamma}_{t,t'}.
\end{equation}
In writing down (\ref{Eq2.15}) we have taken advantage of the fact that
$\Gamma$ is a positive semi-definite matrix, implying
$\lim_{t\to\infty}U_t=E_0$, where $E_0$ denotes the orthogonal projector onto
the null space of $\Gamma$.  Inserting (\ref{Eq2.15}) into (\ref{Eq2.12}), we
find an expression for the stress tensor
\begin{equation}\label{Eq2.16}
  \sigma^{\alpha,\beta}_t =
   \int_{-\infty}^t  d  t'\:
  \left(\frac{d}{dt'}\,G_{t-t'}\right) 
  \sum_{\gamma} T^{\alpha,\gamma}_{t,t'} \, T^{\beta,\gamma}_{t,t'}
\end{equation}
in terms of the time-dependent linear response function
\begin{equation}\label{Eq2.17}
  G_t = 
  \frac{\rho_{0}}{N}\; {\rm Tr}
  \biggl((1-E_0)\exp\biggl\{-\,\frac{2dt}{\zeta a^{2}}\;\Gamma\biggr\}\biggr).
\end{equation}
As a consequence of the simple shear flow (\ref{Eq2.1}) we have
$T_{t,t'}^{\alpha,\beta}=\delta_{\alpha,\beta}+
\delta_{\alpha,x}\delta_{\beta,y}\int^t_{t'}d\tau\kappa(\tau)$.  It follows
from definition (\ref{Eq2.10}) that the static shear viscosity
\begin{equation}\label{Eq2.18}
  \eta({\cal G}) = 
  \frac 1{\rho_{0}}\int_0^{\infty}\! d  t \: G_t = 
  \frac{\zeta a^{2}}{2dN}\:{\rm Tr}\!\left(\frac{1-E_0({\cal G})}
    {\Gamma({\cal G})}\right) 
\end{equation}
is given by the trace of the Moore-Penrose inverse \cite{Al72} of $\Gamma$,
{\it i.e.}\ the inverse of $\Gamma$ restricted to the subspace of non-zero
eigenvalues.

Next we turn to the computation of the incoherent scattering function in the
absence of the shear flow. In this case we have $\kappa =0$ and hence
$T_{t,t'}^{\alpha,\beta}=\delta_{\alpha,\beta}$. Note that ${\bf R}_t - {\bf
  R}_{0}$ is a Gaussian Markov process whose distribution is characterized in
the limit $t_0\to -\infty$ by a vanishing mean and the covariance
\begin{eqnarray}\label{Eq2.19}
  \lefteqn{\lim_{t_0\to -\infty}
  \overline{
    \left(R^{\alpha\vphantom{\beta}}_{t}-R^{\alpha\vphantom{\beta}}_{0}\right)
    \left(R^{\beta}_{t} - R^{\beta}_{0}\right)^{+}}} \nonumber\\
  && \hspace*{2cm} = \frac{2}{\zeta}\: \delta_{\alpha,\beta}
    \int_0^{t} d \tau\:
    \exp\left\{-\frac{d\tau}{\zeta a^2}\:{\Gamma}\right\}.
\end{eqnarray}
To derive (\ref{Eq2.19}) we have used the solution (\ref{Eq2.13}).  Hence the
scattering function is expressed in terms of the diagonal matrix elements of
the $NL\times NL$-matrix
\begin{eqnarray}\label{Eq2.20}
  g_{t} &:=& \frac{1}{\zeta} \int_0^{t} d \tau\:
    \exp\left\{-\frac{d\tau}{\zeta a^2}\:\Gamma\right\} \nonumber\\
    &=& 
    \frac t\zeta \, E_0 +
    \frac{a^2}{d}\frac{1-E_0}{\Gamma}\:
    \left( 1-\exp\!\left\{-\frac{dt}{\zeta a^2}\:\Gamma\right\} \right)
\end{eqnarray}
via
\begin{equation}\label{Eq2.21}
  S_{t}(q|{\cal G})= 
  \frac{1}{NL}
  \sum_{i=1}^N
  \sum_{s=1}^L
  \exp\!\left\{-q^2\:g_t({\cal G}| i,s;i,s)\right\}\,,
\end{equation}
where, again, the dependence on the realization ${\cal G}$ has been made
explicit.
\subsection{Cluster Decomposition}\label{Sec2.4}
Each crosslink realization ${\cal G}$ defines a random labelled graph, which
can be decomposed into its maximal path-wise connected components or clusters
\begin{equation}\label{Eq2.22}
  {\cal G}=\bigcup_{k=1}^{K}{\cal N}_{k}.
\end{equation}
Here, ${\cal N}_{k}$ denotes the $k$-th cluster with $N_k$ polymers out of a
total of $K$ clusters.  The associated connectivity matrix, also called
Kirchhoff matrix or admittance matrix in graph theory, see e.g.\ pp.\
54 in Ref.\ \onlinecite{Bo98}, is
of block-diagonal form
\begin{equation}\label{Eq2.23}
  \Gamma({\cal G})=\bigoplus_{k=1}^K\Gamma({\cal N}_k).
\end{equation}
Therefore the viscosity (\ref{Eq2.18}) is decomposed into contributions from
different clusters according to
\begin{equation}\label{Eq2.24}
  \eta({\cal G}) = \sum_{k=1}^{K} \frac{N_{k}}{N}\:\eta({\cal N}_{k})\,,
\end{equation}
where the individual contribution from cluster ${\cal N}_k$ is defined by
\begin{equation}\label{Eq2.25}
  \eta({\cal N}_{k}) := 
  \frac{\zeta a^{2}}{2dN_{k}}\:{\rm Tr}\!\left(\frac{1-E_0({\cal N}_{k})}
    {\Gamma({\cal N}_{k})}\right) \,.
\end{equation}
In the same way the incoherent scattering function may be decomposed into
contributions from different clusters
\begin{equation}\label{Eq2.26}
  S_t(q|{\cal G})=\sum_{k=1}^{K}
  \frac{N_k}{N}S_t(q|{\cal N}_k),
\end{equation}
where the individual contribution from cluster ${\cal N}_k$ is defined by
\begin{equation}\label{Eq2.27}
  S_t(q|{\cal N}_k):=\frac{1}{N_kL}  
  \sum_{i\in {\cal N}_k}
  \sum_{s=1}^L
  \exp\!\left\{-q^2\:g_t({\cal N}_k|i,s;i,s)\right\}.
\end{equation}
Combining (\ref{Eq2.26}), (\ref{Eq2.27}) and (\ref{Eq2.20}), we
express the scattering function as
\widetext
 \ifpreprintsty      
   \begin{eqnarray} \label{Eq2.29}
     S_t(q|{\cal G})= && \sum_{k=1}^{K}
     \frac{N_k}{N}  
     \exp\!\left\{ -\frac{q^2t}{\zeta N_kL} \right\}
     \sum_{i\in {\cal N}_k}
     \sum_{s=1}^L\frac{1}{N_kL} \nonumber\\
     && \times
     \exp\!\left\{-\frac{q^2a^2}{d}\left[\frac{1-E_0({\cal N}_k)}
         {{\Gamma}({\cal N}_k)}\: \left(
           1-\exp\!\left\{-\frac{dt}{\zeta a^2}\:{\Gamma}({\cal N}_k)\right\}
         \right)\right](i,s;i,s)\right\}\,,
   \end{eqnarray}
 \else               
  \begin{equation} \label{Eq2.29}
    S_t(q|{\cal G})=\sum_{k=1}^{K}
    \frac{N_k}{N}  
    \exp\!\left\{ -\frac{q^2t}{\zeta N_kL} \right\}
    \sum_{i\in {\cal N}_k}
    \sum_{s=1}^L\frac{1}{N_kL} 
    \exp\!\left\{-\frac{q^2a^2}{d}\left[\frac{1-E_0({\cal N}_k)}
        {{\Gamma}({\cal N}_k)}\: \left(
          1-\exp\!\left\{-\frac{dt}{\zeta a^2}\:{\Gamma}({\cal N}_k)\right\}
        \right)\right](i,s;i,s)\right\}\,,
  \end{equation}
 \fi
\narrowtext
where we have used the representation
\begin{equation}\label{Eq2.28}
  E_0({\cal G}|i,s;i',s')=\sum_{k=1}^{K}
  \frac{1}{N_kL}
  \delta_{{\cal N}(i),{\cal N}_k}
  \delta_{{\cal N}_k,{\cal N}(i')}
\end{equation}
of the orthogonal projector $E_{0}$ onto the null space of $\Gamma$. 
Here ${\cal N}(i)$ denotes the cluster containing polymer $i$.
Eq.\ (\ref{Eq2.28}) follows from the fact that the null space of $\Gamma$ is
spanned by the vectors which are constant when restricted to any one
cluster. Physically this reflects that there is no force acting on the
centre of mass of any one cluster. Hence, the number of zero modes of
$\Gamma$ is equal to the total number $K$ of clusters. 

For calculating disorder averages it will be advantageous to reorder
the sums in 
(\ref{Eq2.24}) and (\ref{Eq2.29}) by summing first over all clusters consisting
of a given number $n$ of polymers and subsequently over all ``sizes'' $n$. Thus
we obtain a decomposition of the average
\begin{equation}\label{Eq2.30}
  \left\langle \sum_{k=1}^{K} \frac{N_k}{N} f({\cal N}_{k})\right\rangle
  = \sum_{n=1}^{\infty} n\tau_{n} \left\langle f\right\rangle_{n}
\end{equation}
for an arbitrary real-valued function $f$. In (\ref{Eq2.30}) we have introduced
the average
\begin{equation}\label{Eq2.31}
  \left\langle f\right\rangle_{n} := \frac{1}{\tau_{n}} \left\langle 
    \frac{1}{N}\sum_{k=1}^{K} \delta_{N_{k},n}\, f({\cal N}_{k})\right\rangle
\end{equation}
over all clusters of $n$ polymers, each consisting of $L$ monomers.
Furthermore
\begin{equation}\label{Eq2.32}
  \quad\tau_{n} := 
  \left\langle  \frac{1}{N}\sum_{k=1}^{K} \delta_{N_{k},n}\right\rangle
\end{equation}
represents the average number of clusters of size $n$ per polymer. Note that
up to now no particular crosslink distribution has been specified.
\section{Shear Viscosity}\label{Sec3}
\subsection{Relation to networks of random 
resistors}\label{Sec3.1}
The viscosity (\ref{Eq2.18}) of randomly crosslinked polymers can be quite
generally related to the resistance of a random electrical network. Thanks to
the cluster decomposition (\ref{Eq2.24}) it suffices to consider an arbitrary
connected cluster ${\cal N}_{k}$. To establish this connection we identify a
bond between two neighbouring monomers on the same polymer as a resistor of
magnitude $l^{2}/a^{2}$ and a crosslink between polymers as a resistor of
magnitude $1$.  The resistance measured between any connected pair of vertices
$(i,s)$ and $(i',s')$ will be denoted by ${\cal R}(i,s;i',s')$. For an
arbitrary connected resistor network ${\cal N}_{k}$ the laws of Ohm and
Kirchhoff lead \cite{KlRa93} to a relation between ${\cal R}$ and the
Moore-Penrose inverse of the connectivity matrix
\begin{eqnarray} \label{Eq3.1}
    {\cal R}({\cal N}_k|i,s;i',s')
    &=&\frac{1-E_0({\cal N}_k)}{\Gamma({\cal N}_k)}(i,s;i,s) \nonumber\\
    &&+\;\frac{1-E_0({\cal N}_k)}{\Gamma({\cal N}_k)}(i',s';i',s') \nonumber\\
    && -\;2\frac{1-E_0({\cal N}_k)}{\Gamma({\cal N}_k)}(i,s;i',s')\,.
\end{eqnarray}
The matrix elements of the projector $E_{0}({\cal N}_{k})$ are all equal to
$(N_{k}L)^{-1}$ according to (\ref{Eq2.28}). Hence, when summing (\ref{Eq3.1})
over all $(i,s)$, $(i',s')\in {\cal N}_{k}$, the last term on the right-hand
side equals $-2N_{k}L\, {\rm Tr}\bigl[ E_{0}({\cal N}_{k}) \bigl( 1-
E_{0}({\cal N}_{k})\bigr)/\Gamma({\cal N}_{k})\bigr]$ and thus vanishes. Using
this fact and the definition (\ref{Eq2.25}) of $\eta({\cal N}_{k})$, the
viscosity of a cluster ${\cal N}_{k}$ is seen to be given by a sum over
resistances
\begin{equation}\label{Eq3.2}
  \eta({\cal N}_k) =  
  \frac{\zeta a^{2}}{4d L N_k^2} \sum_{(i,s),(i',s')\in{\cal N}_k}
  {\cal R}({\cal N}_k|i,s;i',s'). 
\end{equation} 
Together with (\ref{Eq2.24}) this constitutes the announced connection between
the viscosity of a randomly crosslinked polymer melt and a random resistor
network.  We would like to emphasize that this connection relies on the special
form (\ref{Eq2.18}) of the viscosity in the Rouse model.
\subsection{Finite-dimensional percolation: scaling
  description}\label{Sec3.2} 
The gelation and the vulcanization transition have been interpreted as a
percolation transition already by Flory and Stockmayer.
\cite{Flo41-42,Sto43-44,Flo53} Even though the classical theories had to be
replaced by modern approaches \cite{Gen78,Sta76} built on the analogy to
critical phenomena, the percolation picture survived. It is generally believed
that the ensemble of macromolecular clusters which are built in the process of
gelation has the same statistical connectivity as an ensemble of clusters which
are constructed in a $d$-dimensional bond percolation process.  This
identification is supported by experiment, e.g.\ the measured cluster-size
distribution of macromolecules is well described by the critical exponents
predicted for bond percolation. \cite{StCoAd82,StAh94}

The most intuitive picture of gelation and vulcanization is related to
continuum percolation, which is expected to be in the same universality class
as random bond percolation. \cite{StCoAd82} In continuum percolation a
set of points is 
randomly distributed in $d$-dimensional space. The points are assumed to be the
centres of spheres with randomly distributed radii. Two points are said to be 
connected if their spheres overlap. Correspondingly, if a crosslinking
agent is added to a dense 
polymer solution or melt, then we expect that pairs of monomers which
are close to each other
will be crosslinked with high probability, provided the reaction is
sufficiently fast as compared to the diffusion time of the polymers.

The well-known characteristics of percolation clusters will allow us
to determine the critical behaviour of the shear viscosity as given by
(\ref{Eq2.24}) and (\ref{Eq2.30}).
For that purpose we first recall some quantities which are
important in the scaling theory of percolation, see e.g.\ Ref.\ 
\onlinecite{StAh94,StCoAd82}. For definiteness we consider bond
percolation on the 
$d$-dimensional cubic lattice ${\Bbb Z}^{d}$.  Nearest-neighbour bonds are
present with probability $p:=c/(2d)$ and absent with probability $1-p$. The
bonds are interpreted as crosslinks, while the lattice points are identified
with monomers.  Here, the probability $p$ is defined such that the
average number of crosslinks is given by
$M=cN$. A percolating cluster first appears at a critical
crosslink concentration $c_{\rm crit}$. The fraction $Q$ of lattice points
belonging to the infinite network serves as an order parameter and vanishes
continuously as the percolation transition is approached: $ Q
\stackrel{c\downarrow c_{\rm crit}}{\sim} (c-c_{\rm crit})^{\beta}$.  The
correlation function ${\cal P}(r)$ is proportional to the probability that
two vertices which belong to the same cluster are a distance $r$ apart.  As the
transition is approached, correlations become increasingly long-ranged
as is indicated by a divergence of the correlation length $\xi$
\begin{equation}\label{Eq3.4}
  \xi^2 
  :=
  \sum_{r\in{\Bbb Z}^d} r^2\,{\cal P}(r)
  \stackrel{c \uparrow c_{{\rm crit}}}{\sim}(c_{{\rm crit}} - c)^{-2\nu}\,.
\end{equation} 
At the critical point the correlation function shows an algebraic decay
\begin{equation}\label{Eq3.5}
  \left.\vphantom{\big(}{\cal P}(r)\right|_{c=c_{\rm crit}}
  \stackrel{r\to\infty}{\sim} r^{-(d-2+\eta)}.
\end{equation}
As far as geometric properties of percolation are concerned, the critical
behaviour is determined by two independent exponents. Here we choose the
exponent $\nu$ of the correlation length and the anomalous dimension $\eta$.
The other exponents may be expressed in terms of those via various scaling
relations, such as
\begin{equation}\label{Eq3.6}
  \beta = \frac{\nu}{2}(d-2+\eta) .
\end{equation}
It is a well-established fact of the scaling description of
percolation that the 
average number of clusters of size $n$ obeys a scaling law
\begin{equation}\label{Eq3.7}
  \tau_n=n^{-\tau}f\bigl((c_{{\rm crit}}-c)\,n^{\sigma}\bigr)  
\end{equation}
with the scaling function $f$ decaying faster than any polynomial for large
arguments and approaching a nonzero constant for small arguments.  The
exponents $\sigma$ and $\tau$ are related to $\eta$ and $\nu$ by
\begin{equation}\label{Eq3.8}
  \tau= 1+\frac{2d}{d+2-\eta},\qquad\quad
  \sigma=\frac{2}{\nu(d+2-\eta)}.
\end{equation}

After these preliminaries we turn to the calculation of the
averaged viscosity. For the sake of simplicity we consider 
polymers with $L=1$, that 
is, a network of Brownian particles. This is relevant for random networks of
macromolecules which are generated by polycondensation, starting from small
units without internal structure. The relation (\ref{Eq3.2}) between the
viscosity of a cluster and the corresponding resistance then simplifies to
\begin{equation}\label{Eq3.3}
  \eta({\cal N}_k) =  
  \frac{\zeta a^{2}}{4d N_k^2} \sum_{i,i'\in{\cal N}_k}
  {\cal R}({\cal N}_k|i;i'). 
\end{equation}
We remark that the average $\left\langle \eta\right\rangle_n$, which
was defined in 
(\ref{Eq2.31}), is an expectation value over all lattice animals of size $n$
and $N \tau_n$ refers to the average number of percolation clusters of size
$n$.

We make the basic assumption that $\left\langle\eta\right\rangle_n$ which,
according to (\ref{Eq2.31}), only tests clusters of a given finite size $n$,
does not acquire any irregularity as the critical point is approached and that
$\left.\left\langle\eta\right\rangle_n\right|_{c=c_{\rm crit}}\sim n^b$ with
some yet unknown exponent $b$.  We will see below in (\ref{Eq3.21}) that this
holds true with $b=1/2$ in the special case where the crosslinks are
distributed according to mean-field percolation.
Furthermore, we remark that this assumption may be circumvented in a more
sophisticated approach, as will be explained below.

Using the properties of the scaling function $f$ and (\ref{Eq3.8}),
Eqs.\ (\ref{Eq2.24}) and (\ref{Eq2.30}) yield for the asymptotic
behaviour of the viscosity
\begin{eqnarray}\label{Eq3.9}
  \langle\eta\rangle 
  & \stackrel{c\uparrow c_{\rm crit}}{\sim} &
  \sum_{n=1}^\infty n\tau_n 
  \left(\left.\vphantom{\big(}
      \left\langle\eta\right\rangle_n\right|_{c=c_{\rm crit}} 
  \right)
  \sim\sum_{n=1}^\infty n\tau_nn^b \nonumber\\
  & \stackrel{c\uparrow c_{\rm crit}}{\sim} &
  (c_{\rm crit}-c)^{-A(b)},
\end{eqnarray} 
with a critical exponent $A(b):=\frac 12b\nu(d+2-\eta)-\frac 12\nu(d-2+\eta)$.
Thus, the remaining task is the derivation of the exponent $b$.

Since the critical behaviour of the viscosity is believed not to be determined
solely by large-scale geometrical properties, {\it i.e.}\ by the
exponents $\eta$ and 
$\nu$, the appearance of a new exponent is expected.  From (\ref{Eq3.3}) and
(\ref{Eq2.31}) the average over clusters of size $n$ is written in the form
\begin{eqnarray}\label{Eq3.10}
  \left\langle \eta \right\rangle_n = 
  \frac{\zeta a^{2}}{4dn^2\tau_{n}}  
  \Bigg\langle 
    \frac{1}{N}\sum_{k=1}^K \delta_{N_k,n}
    \sum_{i,i'=1}^N &&
    \delta_{{\cal N}(i),{\cal N}_k}
    \delta_{{\cal N}_k,{\cal N}(i')} \nonumber\\
    && \times\; {\cal R}(i;i')  \Bigg\rangle,
\end{eqnarray}
which yields
\begin{eqnarray}\label{Eq3.11}
    \frac{4d}{\zeta a^{2}}\;\sum_{n=2}^{\infty} n^2
    \tau_n\left\langle\eta\right\rangle_n &= &
    \left\langle
      \frac{1}{N}\sum_{i,i'=1}^N \delta_{{\cal N}(i),{\cal N}(i')}
      {\cal R}(i;i')
    \right\rangle \nonumber\\
    &=&
    \left\langle \sum_{i'\in{\cal N}(i)} {\cal R}(i;i') \right\rangle 
    \nonumber \\ 
    &\stackrel{c \uparrow c_{{\rm crit}}}{\sim} &
    (c_{{\rm crit}} - c)^{-(2-\eta)\nu-\phi}.
\end{eqnarray}
To obtain the second equality in (\ref{Eq3.11}) we used enumeration invariance
of the system, and the asymptotic behaviour for $c \uparrow c_{{\rm crit}}$ is
obtained from Eq. (2.45) in Ref.\ \onlinecite{HaLu87}.  The exponent
$\phi$, which was first 
introduced in the context of random resistor networks, governs
\cite{LuWa85,HaLu87,StJaOe99} the growth of
the resistance ${\cal R}(r)$ between two points on the incipient spanning
cluster, which are a large spatial distance $r$ apart: ${\cal R}(r)\sim
r^{\phi/\nu}$.  Referred to as the crossover
resistance exponent, \cite{HaLu87,StJaOe99} $\phi$ is related \cite{StAh94} to
the spectral dimension, according to $\phi= \nu d_{f}\bigl((2/d_{s})-1\bigr)$,
where $d_{f}=d-\beta/\nu$ is the Hausdorff-Besicovitch dimension of the
incipient spanning cluster.  Using again the scaling assumptions (\ref{Eq3.7})
and $\left.\left\langle\eta\right\rangle_n\right|_{c=c_{\rm crit}}\sim n^b$,
Eq.\ (\ref{Eq3.11}) amounts to
\begin{equation}\label{Eq3.12}
  \sum_{n=1}^\infty n\tau_n n^{b+1}
  \stackrel{c \uparrow c_{{\rm crit}}}{\sim} 
  (c_{{\rm crit}} - c)^{-(2-\eta)\nu-\phi},
\end{equation}
which, upon comparison with (\ref{Eq3.9}) gives $A(b+1)=(2-\eta)\nu+\phi$ and,
after a little algebra, the relation 
\begin{equation}
  \label{bexp}
  b=\sigma\phi = (2/d_{s}) - 1\,.
\end{equation}
Thus we have derived the scaling relation
\begin{equation}\label{Eq3.13}
  k=\phi-\beta
\end{equation}
for the viscosity exponent $k$.

As already mentioned, the above scaling assumption is encoded in a more general
scaling relation. In Refs.\ \onlinecite{HaLu87} and
\onlinecite{StJaOe99} the authors 
discuss the generating function
\begin{equation}\label{Eq3.14}
  Z(\lambda,\omega):= 
  \left\langle 
    \exp\!\left\{-\frac{\lambda^{2}}{2}
      \left(\frac
        1{\Gamma+i\omega}\right)_{j,j}\right\}\right\rangle
\end{equation}
of the distribution of the resolvent of $\Gamma$. Amongst other things
they show 
by means of a renormalization-group analysis up to second order in
$\varepsilon=6-d$ the validity of the scaling relation
\begin{eqnarray}\label{Eq3.15}
  \lefteqn{Z(\lambda,\omega)
    \stackrel{c\uparrow c_{\rm{crit}}}{\sim}
    (c_{\rm{crit}}-c)^{\beta} }\nonumber\\
  && \hspace*{1.8cm} \times\; z\left((c_{\rm{crit}}-c)^{-\phi}\lambda^2,
    (c_{\rm{crit}}-c)^{1/\sigma}\omega\right)  
\end{eqnarray}
with an appropriate scaling function $z$.  We use the fact that the system is
translationally invariant and relate the viscosity to $Z$ via
\begin{eqnarray}\label{Eq3.16}
  \left\langle \eta\right\rangle
  &=& \;
  \frac{\zeta\,a^2}{2d}\left\langle
    \left(\frac{1-E_0}{\Gamma}\right)_{j,j}\right\rangle  
  \nonumber\\ 
  &=& -\;\frac{\zeta\,a^2}{d}\lim_{\omega\to 0} 
     \left.\frac{\partial}{\partial\lambda^2}\right|_{\lambda=0} 
      Z(\lambda,\omega)
\end{eqnarray}
To derive the second equality we have employed the expansion
\begin{equation}\label{Eq3.17}
  \frac{1}{\Gamma+i\omega} = - \frac i\omega E_0 + \frac{1-E_0}\Gamma 
   + {\cal O}(\omega), \quad \omega \to 0, 
\end{equation}
in (\ref{Eq3.14}). From (\ref{Eq3.16}) the result (\ref{Eq3.13}) is readily
re-derived.
\subsection{Mean-field percolation: exact results}\label{Sec3.3}
To go beyond scaling arguments and to compute the viscosity for all
crosslink concentrations,
we have to resort to simpler cluster distributions than the one
given by bond percolation in finite dimensional space.  We only consider
the simplest distribution which is equivalent to mean-field percolation, as far
as static quantities are concerned. All pairs of monomers are equally likely to
be crosslinked and no correlations between crosslinks are taken into account.
More precisely, for a function $f$ depending on the realization ${\cal
  G}=\{i_e, s_{e};i_e', s'_{e}\}_{e=1}^{M}$ of the network we define the
average by
\begin{eqnarray}\label{Eq3.18}
  \langle f\rangle := 
  \lim_{N,M\to\infty \atop  M/N =c} 
  \left( \prod_{e=1}^{M}
  \frac{1}{(NL)^{2}} \right. && \left. \sum_{i_e,i'_e=1}^{N}
  \sum_{s_e,s'_e=1}^{L}\right)
  \nonumber\\
  && \times\; f(\{i_e, s_{e};i_e', s'_{e}\}).
\end{eqnarray}
The static shear viscosity can be computed exactly for this simple
distribution, making use of results in the mathematical literature on random
graphs. The calculations are easiest for a network of Brownian particles, that
is, polymers consisting of just one monomer each ($L=1$). The generalization to
more complex molecular units with arbitrary $L$---not necessarily
linear---will be shown to be 
straightforward. It will not change the critical behaviour as compared to the
case $L=1$.
\subsubsection{Brownian particles: $L=1$}
The statistical properties of clusters generated according to the above
distribution, have been studied extensively in the theory of random graphs, as
developed in Ref. \onlinecite{ErRe60}. Strictly speaking, the ensemble
of random graphs considered in Ref.\ \onlinecite{ErRe60} comprises
random labelled graphs consisting of $N$ 
vertices, $M=cN$ edges and each of the $\left({\left(N \atop
      2\right)}\atop M\right)$
possible graphs is equally likely to occur, {\it i.e.} double edges or
self-loops are suppressed.  Obviously this ensemble differs from the one
considered in the expectation (\ref{Eq3.18}), but one expects both to have the
same properties in the macroscopic limit.
This may be understood by comparing
the number ${\cal O}(N)$ of possibilities to realize a self-loop or double edge
in ${\cal G}$ with the number ${\cal O}(N^2)$ of possibilities to choose a
different monomer.

In the macroscopic limit there are no clusters of macroscopic size for
$c<c_{{\rm crit}}:=\frac{1}{2}$, and all monomers belong to tree clusters
without loops, see Thms.~5d,e in Ref.\ \onlinecite{ErRe60}.  Moreover,
according to Eq.\  (2.18) in Ref.\ \onlinecite{ErRe60}, one has
\begin{equation}\label{Eq3.19}
  \tau_{n}=\frac{n^{n-2}}{2c\,n!} \, (2c\,{\rm e}^{-2c})^n 
\end{equation}
for the average number of trees of size $n$ per polymer.  Since each cluster
${\cal N}_k$ of size $n=N_k$ is expected to be realized as exactly one of the
$n^{n-2}$ labelled trees ${\cal T}_n$ of size $n$, one has $\sum_{{\cal
    T}_n}\delta_{{\cal N}_k,{\cal T}_n}=\delta_{N_k,n}$, and the expectation
(\ref{Eq2.31}) over clusters of size $n$ is expressed as an expectation over
trees
\begin{equation}\label{Eq3.20}
    \left\langle f\right\rangle_{n}= 
    \sum_{{\cal T}_{n}} f({\cal T}_{n})
    \frac{1}{\tau_n}
    \left\langle \frac{1}{N}\sum_{k=1}^{K} \delta_{{\cal N}_{k},{\cal
          T}_n} \right\rangle
    =\frac{1}{n^{n-2}}\sum_{{\cal T}_{n}} f({\cal T}_{n}).
\end{equation}
The last step is understood as follows: The assumption that ${\cal N}_k$ is a
tree of size $n$ fixes the number of vertices {\it and} edges in ${\cal N}_k$.
Hence, due to the independence of the crosslinks in (\ref{Eq3.18}), the number
of graphs, which may be realized within this assumption, depends only on the
size $n$, and all trees ${\cal T}_n$ can be proven to occur equally likely as a
realization of ${\cal N}_k$. Accordingly, the probability
$\tau_n^{-1}\left\langle N^{-1}\sum_{k=1}^{K} \delta_{{\cal
      N}_{k},{\cal T}_n}\right\rangle$ 
for a cluster ${\cal N}_k$ of size $n$ to be realized as the tree ${\cal T}_n$
is equal to $n^{2-n}$.

Note further that the resistance between two vertices $i$ and $i'$ in a tree
simplifies considerably, because there is a unique path connecting $i$ and
$i'$, implying that all resistors are in series.  Therefore the resistance
${\cal R}(i;i')$ is equal to the number of crosslinks connecting $i$ and $i'$.
Hence, we refer to Thms.~1,2 in Ref.\ \onlinecite{MeMo70} for computing
\begin{eqnarray}\label{Eq3.21}
  \left\langle {\cal R}(i;i')\right\rangle_{n} &=& 
  (n-2)!\,\sum_{\nu=2}^{n} \frac{n^{1-\nu}\,\nu(\nu -1)}{(n-\nu)!} \nonumber\\
  &\stackrel{n\to\infty}{\sim} &\sqrt{n\pi/2}\,, \qquad\qquad i\neq i'\,.
\end{eqnarray}
Obviously, one has $\left\langle {\cal R}(i,i)\right\rangle_n =0$.  
Now (\ref{Eq3.2}) and (\ref{Eq3.21}) imply
\begin{equation}
  \label{etan1}
  \left\langle \eta\right\rangle_{n} = \frac{\zeta a^{2}}{4d}\:
  (n-1)!\,\sum_{\nu=2}^{n} \frac{n^{-\nu}\,\nu(\nu -1)}{(n-\nu)!} \,.
\end{equation}
Combining (\ref{Eq2.24}), (\ref{Eq2.30}) and
(\ref{etan1}), and expanding the exponential in
(\ref{Eq3.19}), one obtains
\begin{eqnarray} 
    \left\langle\eta\right\rangle
    &=&\frac{\zeta\,a^2}{8dc}
    \sum_{n=2}^{\infty}
    \sum_{l=0}^{\infty}
    \sum_{\nu=2}^{n}
    (-1)^l\;
    \frac{(2c)^{n+l}n^{n+l-\nu-2}\nu(\nu-1)}
    {l!(n-\nu)!} \nonumber\\
    &=&\frac{\zeta\,a^2}{8dc}
    \sum_{j=2}^{\infty}
    (2c)^{j}
    \sum_{l=0}^{j-2}
    \sum_{\nu=2}^{j-l}  
    (-1)^l\frac{(j-l)^{j-\nu-2}\nu(\nu-1)}{l!(j-l-\nu)!} \nonumber\\
    &=&\frac{\zeta\,a^2}{8dc}
    \sum_{j=2}^{\infty}
    (2c)^{j}
    \sum_{\nu=2}^{j}\frac{\nu(\nu-1)}{(j-\nu)!} \nonumber\\
    && \qquad\times\;\sum_{l=0}^{j-\nu}
    (-1)^l
    (j-l)^{j-\nu-2}\left({j-\nu \atop l}\right)\,. \label{Eq3.22}
\end{eqnarray} 
To sum up this expression in closed form, we note that
\begin{eqnarray}\label{Eq3.23}
  \sum_{l=0}^m
  (-1)^l\, &&
  (\alpha+m-l)^{m-2}
  \left({m \atop l}\right) \nonumber\\
  && = \left\{
  \begin{array}{cl}
    \:\alpha^{-2}, &\quad {\rm for}\: m=0,\\
    \:- \bigl(\alpha(\alpha+1)\bigl)^{-1},&\quad{\rm for}\:m=1,\\
    \:0,&\quad{\rm otherwise}.
  \end{array}\right.
\end{eqnarray}
For $m=0,1$ this is verified by inspection and for $m\ge2$ by differentiating
twice Formula 0.154.5 in Ref.\ \onlinecite{GrRh80} with respect to
$\alpha$.  Hence the 
result of the $\nu$ and $l$ sums in (\ref{Eq3.22}) is equal to $1/j$ and we
obtain
\begin{equation}\label{Eq3.24}  
    \left\langle\eta\right\rangle = 
    \frac{\zeta\, a^2}{8dc}
    \left[ \ln\!\left(\frac{1}{1-2c}\right) - 2c\right]
    , \quad \text{for $L=1$},
\end{equation}
for the averaged viscosity for all $0<c<\frac{1}{2}$.  The viscosity exhibits a
logarithmic divergence at the critical concentration $c_{{\rm crit}} =
\frac{1}{2}$ corresponding to the critical exponent $k=0$. This is in
accordance with the more general result (\ref{Eq3.13}) because for $d\ge 6$ one
has \cite{StAh94} $\beta=1=\phi$. The behaviour (\ref{Eq3.24}) of the
averaged viscosity is displayed in Figure~\ref{fig1} in the
Introduction.

\subsubsection{Network of polymers: General $L$}
\begin{figure}[t]
  \center \leavevmode
  \epsfxsize=20.5pc
  \epsffile{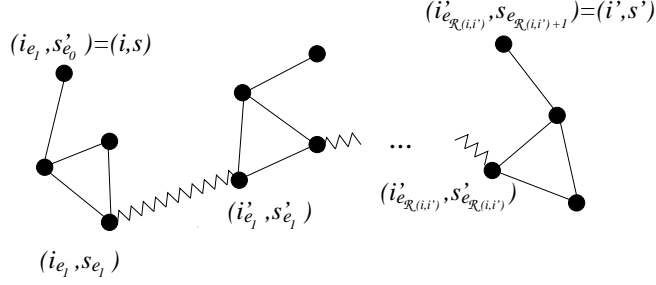} 
  \medskip
  \caption{\label{fig2}Contributions to the total resistance in Eq.\
    (\protect\ref{Eq3.26}). Crosslinks are depicted as zig-zag lines,
    intra-polymer bonds as straight lines.}
\end{figure}

The result (\ref{Eq3.24}) is readily generalized to networks of crosslinked
polymers, all having the same number $L\ge 1$ of monomers and the same
geometric structure. The only difference is that now there are two different
kinds of resistors with magnitudes $l^{2}/a^{2}$ and $1$, corresponding to
intra-polymer bonds and crosslinks, respectively. Due to the fact that the
monomer labels $s$ are distributed independently from the chain labels $i$ in
the expectation (\ref{Eq3.18}), it follows by the same arguments as above that
in the macroscopic limit all polymers are connected within tree clusters
$\{i_{e},i'_{e}\}$. Thus the average (\ref{Eq3.20}) over all possible
configurations of clusters consisting of $n$ polymers generalizes to
\begin{equation}\label{Eq3.25}
  \left\langle f\right\rangle_{n} =
  \left(
    \prod_{e=1}^{n-1}
    \frac{1}{L^2}
    \sum_{s_e,s'_e=1}^L
  \right)
  \frac{1}{n^{n-2}}
  \sum_{{\cal T}_{n}} 
  f(\{i_e,s_e;i_e',s_e'\})\,.
\end{equation}
In (\ref{Eq3.25}) the average is taken over equally probable trees ${\cal
  T}_{n}=\{i_{e},i'_{e}\}_{e=1}^{n}$ and over independently chosen, equally
probable monomer labels.  Given $i,i'\in{\cal T}_n$, the resistance ${\cal
  R}(i,s;i',s')$ splits up into two additive contributions, see also
Fig.\ \ref{fig2}. 
The first is the
crosslink part ${\cal R}(i;i')$, which is equal to the number of inter-polymer
crosslinks, that is resistors of magnitude one, on the unique path on the tree
${\cal T}_{n}$ from $i$ to $i'$. Let us denote these crosslinks by
$\{i_{e_k},s_{e_k};i'_{e_k},s'_{e_k}\}_{k=1}^{{\cal R}(i,i')}$.  The second
stems from intra-polymer resistors of magnitude $l^2/a^2$. 
Thus, we write
\begin{equation}\label{Eq3.26}
    {\cal R}(i,s;i',s')=
    {\cal R}(i,i') +\frac{l^2}{a^2} \sum_{k=0}^{{\cal R}(i,i')}
    {\cal Q}(s'_{e_{k}},s_{e_{k+1}}) \,,
\end{equation}
where $s'_{e_{0}}:=s$, $s_{e_{{\cal R}(i,i') +1}}:=s'$ and $(l^{2}/a^{2}){\cal
  Q}(\sigma,\sigma')$ is the resistance measured between monomer $\sigma$ and
monomer $\sigma'$ on the same polymer.
It follows that
\begin{equation}\label{Eq3.27}
  \frac{1}{L^{2}}\, \sum_{s,s'=1}^{L} \left\langle \sum_{k=0}^{{\cal R}(i,i')}
    {\cal Q}(s'_{e_{k}},s_{e_{k+1}})\right\rangle_{n}  \!\!=
  \bigl( 1+ \left\langle {\cal R}(i,i')\right\rangle_n \bigr) \, {\cal
    Q}_{L} \,. 
\end{equation}
The latter holds due to the fact that monomer and polymer labels are
distributed independently in the average (\ref{Eq3.25}), and we have introduced
the averaged single-polymer resistance
\begin{equation}\label{Eq3.28}
  {\cal Q}_{L} := \frac{1}{L^{2}}\, \sum_{\sigma,\sigma'=1}^{L} {\cal
    Q}(\sigma,\sigma')\,. 
\end{equation}
Upon inserting (\ref{Eq3.26}) into (\ref{Eq3.2}), these two relations lead to
\begin{equation}\label{Eq3.29}
  \left\langle \eta\right\rangle_n= L
  \left(1 +\frac{l^{2}}{a^{2}}{\cal Q}_{L}\right)\;
  \left\langle \eta\right\rangle_{n} \bigr|_{L=1} 
  +\frac{\zeta l^{2}}{4d}\; L{\cal Q}_{L}\,.
\end{equation}
Thus, the averaged viscosity for general $L\ge 1$ follows from
(\ref{Eq2.30}), (\ref{Eq3.29}), the 
averaged viscosity (\ref{Eq3.24}) of the corresponding Brownian-particle case
and the relation $\sum_{n=1}^{\infty}n\tau_n=1$, which is valid for all
$0<c<\frac{1}{2}$.  This gives rise to the exact result
\begin{eqnarray}\label{Eq3.30}
  \left\langle \eta\right\rangle  & = & 
  \frac{\zeta a^2}{8cd} L \left( 1 + \frac{l^{2}}{a^{2}}\;{\cal Q}_{L} \right)
  \left[ \ln\!\left(\frac{1}{1-2c}\right) - 2c \right] \nonumber\\
  &&+  \frac{\zeta l^{2}}{4d}\; L{\cal Q}_{L}
\end{eqnarray}
for the averaged viscosity of a crosslinked polymer melt on the sol side. As in
the Brownian-particle case, it exhibits a logarithmic divergence at the
critical crosslink concentration $c_{{\rm crit}}=\frac{1}{2}$.  The result
(\ref{Eq3.30}) is universal in the sense that the details of the model only
affect the pre-factor of the critical divergence.  The pre-factor depends on
the persistence length $l$, the ``extension'' of the crosslinks $a$, as well as
on the number $L$ of monomers of a polymer and on the geometric structure of a
polymer through the averaged single-polymer resistance ${\cal
  Q}_{L}$. Let us mention three concrete examples for the latter.
First, in the
special case where the polymers are chain molecules, the averaged
single-polymer resistance is given by ${\cal Q}_{L} = L^{-2}
\sum_{\sigma,\sigma'=1}^{L} |\sigma - \sigma'| = (L^{2}-1)/(3L)$. Accordingly,
for long chains the viscosity is proportional to $L^2$, as it should be within
a Rouse-type model. \cite{DoEd85}  Second, for ring
polymers ${\cal Q}_{L} = (L^{2}-1)/(6L)$ is half as big as
for linear chains. Third, for star polymers one obtains 
${\cal Q}_{L} = 2 (L-1)^{2}/L^{2}$ which implies a linear growth 
of $\left\langle \eta\right\rangle$ in $L$ for large $L$.
In all three cases the scaling of the prefactor with $L$ in (\ref{Eq3.30}) 
is not altered when passing to the limit $a\downarrow 0$ of hard
crosslinks.

\subsection{Alternative derivation with replicas}\label{Sec3.4}

In this section we rederive the result (\ref{Eq3.24}) for the disorder
average of the viscosity of a network of Brownian particles by means
of the replica trick. We consider the 
mean-field distribution of 
crosslinks, where each crosslink $(i,i')$, $1\le i<i'\le N$, occurs
independently with probability $2c/N$. In many
respects this ensemble is equivalent in the macroscopic limit to
the one with a fixed number of crosslinks, which underlies
(\ref{Eq3.24}) and was defined by the
expectation (\ref{Eq3.18}) for $L=1$.
In particular, $c_{\rm{crit}}=\frac{1}{2}$ and the expression
(\ref{Eq3.19}) for the average number $\tau_{n}$ of trees of size $n$
per polymer remain valid, as can be inferred from Thms.\ V.7 and V.5
in Ref.~\onlinecite{Bol85}. Additionally, Eq.\ (\ref{Eq3.20}) continues
to hold, because the crosslinks are distributed independently and with
equal probabilities. Therefore the averaged viscosity is given by
(\ref{Eq3.24}) for both ensembles. 

In this subsection we focus again on the non-percolating regime
$c< c_{\rm{crit}}=\frac{1}{2}$. 
To leading order in $N\gg 1$ the ensemble, where the crosslinks are
distributed with probability $2c/N$, is characterized by the
characteristic function 
\begin{equation}\label{eq:re2}
  \left\langle{\rm e}^{{\rm Tr}(\Gamma A)}\right\rangle=
  \exp\left\{\frac{c}{N}\sum_{i,i'=1}^N
    {\rm e}^{A_{i,i}+A_{i',i'}-A_{i,i'}-A_{i',i}}-cN\right\}
\end{equation} 
of the connectivity matrix $\Gamma$. Here, $A$ denotes an arbitrary
$N\times N$-matrix. Note that, contrary to our previous conventions,
in this subsection we do not include the limit $N\to\infty$ in the
average $\langle\bullet\rangle.$ The ensemble (\ref{eq:re2}) of random
matrices is closely related to that in Ref.\
\onlinecite{RoBr88}, where the density of states of $\Gamma$ was
investigated with the replica trick; see also 
Refs.\ \onlinecite{BiMo99} and \onlinecite{CaGiPa99} for recent works
devoted to this issue. 

The following computation of $\langle\eta\rangle$ with replicas has a
lot in common with that of the density of states in
Ref.\ \onlinecite{RoBr88}. Therefore we have chosen a less detailed
exposition in this subsection and only concentrate on the main
steps. Compared to Ref.\ \onlinecite{RoBr88} there is one big
difference, however. For the purpose of $\langle\eta\rangle$ one can
solve the resulting saddle-point equation exactly by analytic means.

The basic quantity in the replica calculation is the averaged
resolvent or Green function
\begin{eqnarray}\label{eq:re3}
  G(\Omega)&:&=\lim_{N\to\infty}
    \frac{1}{2N}
    \left\langle{\rm Tr}\left(\frac{1}{\Gamma+\Omega}\right)\right\rangle 
    \nonumber\\
    &&=\lim_{N\to\infty}
    \frac{1}{2N}\left(
\left\langle{\rm Tr}\left(\frac{1-E_0}{\Gamma+\Omega}\right)\right\rangle
      +\frac{1}{\Omega}
      \bigl\langle\rm{Tr}(E_0)\bigl\rangle\right) \nonumber\\
\end{eqnarray}
of $\Gamma$, in terms of which the viscosity (\ref{Eq2.18}) is written as 
\begin{eqnarray}\label{eq:re4}
  \lim_{N\to\infty}\left\langle\eta\right\rangle & = &
  \frac{\zeta a^2}{d}
  \lim_{\Omega\to 0}
  \left(G(\Omega)-\lim_{N\to\infty}
    \frac{\left\langle{\rm Tr}(E_0)\right\rangle}{2N\Omega}\right)
  \nonumber\\ 
  & = & \frac{\zeta a^2}{d}
  \lim_{\Omega\to 0}
  \left(G(\Omega) -  \frac{1-c}{2\Omega}\right)\,.
\end{eqnarray}
To get the second equality in (\ref{eq:re4}), we have used that
the dimensionality of the null space of $\Gamma$ is equal to the total
number of clusters and refer to Thms.\ V.7(ii) and V.12 in
Ref.\ \onlinecite{Bol85}. In the following we compute $G(\Omega)$ with
replicas and show that it has a singular contribution for $\Omega\to
0$ which cancels the second term in (\ref{eq:re4}). The regular part of
$G(\Omega)$ determines the viscosity.

Introducing the generating function
\begin{eqnarray}\label{eq:re5}
  Z_N(\Omega):=&&\int_{{\Bbb R}^N} 
  \left(\prod_{i=1}^N \frac{ d {x}_i}{\sqrt{2\pi}} \right) \nonumber\\
  &&\!\times\exp{\left(-\frac{1}{2}\sum_{i,i'=1}^N {x}_i{x}_{i'}
      (\Gamma_{i,i'}+\Omega\:\delta_{i,i'})\right)}, 
\end{eqnarray}
the resolvent (\ref{eq:re3}) is expressed as
\begin{equation}\label{eq:re6}
  G(\Omega)=-\lim_{N\to\infty}\frac{1}{N}
  \left\langle\frac{\partial\ln Z_N(\Omega)}{\partial\Omega}\right\rangle\,.
\end{equation}
Now the standard replica technique is applied to perform the average of
the logarithm in (\ref{eq:re6}) via 
$\ln Z_N=\lim_{n\to 0} (Z_N^n-1)/n$.
In so doing we introduce hatted vectors $\hat{{x}}:=({x}^1,{x}^2,\ldots,{x}^n)$
for $n$-times replicated variables. Then, specifying
$A_{i,i'}:=-(\hat{x}_i\cdot\hat{x}_{i'})/2$ in the characteristic
function 
(\ref{eq:re2}), we get for the disorder average
\widetext
\begin{equation}\label{eq:re7}
  \left\langle{\bigl(Z_N(\Omega)\bigr)^n}\right\rangle =
  \int_{{\Bbb R}^{Nn}}
  \left(\prod_{i=1}^N
    \frac{ d \hat{x}_i}{(2\pi)^{n/2}}\right) \,
  \exp{\left(-\frac{\Omega}{2}\sum_{i=1}^N\hat{{x}}_i^2 
      +\frac{c}{N}\sum_{i,i'=1}^Ne^{-(\hat{{x}}_i-\hat{{x}}_{i'})^2/2}
      -cN\right)}\,.
\end{equation}
Next we introduce a delta-correlated Gaussian random field
$\psi:{\Bbb R}^n\to{\Bbb R}$ with zero mean which serves to
decouple the double sum over exponentials in the exponent of
(\ref{eq:re7}). This leads to the functional-integral representation
\begin{equation}\label{Eq3.46}
  \left\langle{\bigl(Z_N(\Omega)\bigr)^n}\right\rangle=
  \int{\cal D}\psi
  \exp\big(-N\:F(\Omega,n,\psi)\bigl),
\end{equation}
where
\ifpreprintsty  
  \begin{eqnarray}\label{eq:re8}
    F(\Omega,n,\psi):=c+
    \frac{1}{2}\int_{{\Bbb R}^n} d  {\hat{x}} &&
    \:\bigl(\psi({\hat{{x}}})\bigr)^2
    -\ln\!\left[
      \int_{{\Bbb R}^n}\frac{ d  {\hat{x}}}{(2\pi)^{n/2}}
      \:{\rm e}^{-\Omega\, {\hat{{x}}}^2/2} \right. \nonumber\\
      &&\times \left.\exp\left( \sqrt{2c}\left(\frac{2}{\pi}\right)^{n/4}
        \int_{{\Bbb R}^n} d {\hat{y}}\:
        \psi(\hat{{y}})\,
        {\rm e}^{-({\hat{{y}}}-{\hat{{x}}})^2}\right)
    \right].
  \end{eqnarray}
\else           
  \begin{equation}\label{eq:re8}
    F(\Omega,n,\psi):=c+
    \frac{1}{2}\int_{{\Bbb R}^n} d  {\hat{x}}
    \:\bigl(\psi({\hat{{x}}})\bigr)^2
    -\ln\!\left[
      \int_{{\Bbb R}^n}\frac{ d  {\hat{x}}}{(2\pi)^{n/2}}
      \:{\rm e}^{-\Omega\, {\hat{{x}}}^2/2}
      \exp\left( \sqrt{2c}\left(\frac{2}{\pi}\right)^{n/4}
        \int_{{\Bbb R}^n} d {\hat{y}}\:
        \psi(\hat{{y}})\,
        {\rm e}^{-({\hat{{y}}}-{\hat{{x}}})^2}\right)
    \right].
  \end{equation}
\fi
\narrowtext
The saddle-point method is now applied to evaluate (\ref{Eq3.46}) in
the limit $N\!\to\!\infty$, yielding
$\left\langle{Z_N(\Omega)^n}\right\rangle
\stackrel{N\to\infty}{\sim}\exp\bigl(-NF(\Omega,n,\tilde{\psi})\bigr)$
where $\tilde{\psi}$ is the unique solution
of the saddle-point equation
\begin{equation}\label{Eq3.49}
  \frac{\delta}{\delta{\psi}} 
  F(\Omega,n,\psi)\Big|_{\psi=\tilde{\psi}}=0.
\end{equation}
Note that existence and uniqueness of $\tilde{\psi}$ are guaranteed, since
for
crosslink concentrations $0<c<\frac{1}{2}$ it is straightforward to prove that
$F(\Omega,n,\psi)$ is strictly convex in the argument $\psi$.
For the following it will be more advantageous to work with 
the Gaussian transform 
\begin{equation}\label{Eq3.48}
  \phi(\hat{{x}})  
  :=\sqrt{2c}\:\left(\frac{2}{\pi}\right)^{n/4}
  \int_{{\Bbb R}^n} d {\hat{y}}\:
  \tilde{\psi}(\hat{{y}})\:
  {\rm e}^{-({\hat{{y}}}-\hat{{x}})^2}
\end{equation}
of the solution $\tilde{\psi}$,
in terms of which the saddle-point equation (\ref{Eq3.49}) 
is written as 
\begin{equation}\label{eq:re9}
  \phi(\hat{{x}})=2c\;\:
  \frac{\displaystyle
    \int_{{\Bbb R}^n} d {\hat{y}}
    \:{\rm e}^{-\Omega\,{\hat{{y}}}^2/2 + \phi(\hat{{y}})}\:
    {\rm e}^{-({\hat{x}}-{\hat{y}})^2/2}}
  {\displaystyle \int_{{\Bbb R}^n} d {\hat{y}}
    \:{\rm e}^{-\Omega\,{\hat{{y}}}^2/2 + \phi(\hat{{y}})}}\,,
\end{equation}
and the resolvent (\ref{eq:re6}) reads
\begin{equation}\label{eq:re10}
  G(\Omega)=
  \lim_{n\to 0}\:
  \frac{1}{2n}\;\:
  \frac{\displaystyle
    \int_{{\Bbb R}^n} d {\hat{y}}\: \hat{y}^{2}
    \:{\rm e}^{-\Omega\,{\hat{{y}}}^2/2 + \phi(\hat{{y}})}}
    {\displaystyle  \int_{{\Bbb R}^n} d {\hat{y}}
      \:{\rm e}^{-\Omega\,{\hat{{y}}}^2/2 + \phi(\hat{{y}})}}\,.
\end{equation}
Note that (\ref{eq:re9}) implies the normalization condition
\begin{equation}\label{Eq3.52}
  \int_{{\Bbb R}^{n}} d\hat{x}\; \phi({\hat{x}})
  =2c\:(2\pi)^{n/2}\,.
\end{equation}

In order to compute the viscosity (\ref{eq:re4}),
the equations (\ref{eq:re9}) and (\ref{eq:re10}) have to be solved
for small $\Omega$ after a continuation to $n=0$.
We restrict ourselves to solutions $\phi$ of (\ref{eq:re9}), which
preserve rotational invariance in replica space and make the
replica-symmetric ansatz  
\begin{equation}\label{Eq3.53}
  \phi({\hat {x}})=\varphi^{\Omega}_n
  \left(\sqrt{\Omega}\:|{{\hat {x}}}|\right)\,.
\end{equation}
For later purposes the dependence on parameters has been made explicit.
Introducing $n$-dimensional spherical coordinates,
Eq.\ (\ref{eq:re10}) transforms into
\begin{equation}\label{Eq3.55}
  G(\Omega)=
  \lim_{n\to 0}
  \frac{1}{2n\Omega}\:\;
  \frac{\displaystyle
    \int_0^{\infty} d  \rho\:\rho^{n+1}
    {\rm e}^{-\rho^2/2 +
      \varphi^{\Omega}_n(\rho)}}
      {\displaystyle \int_0^{\infty} d  \rho\:\rho^{n-1}
    {\rm e}^{-\rho^2/2 +
      \varphi^{\Omega}_n(\rho)}}\,.
\end{equation}
To proceed with Eq.\ (\ref{eq:re9})
we employ an integral representation of the exponential of the
$n$-dimensional Laplacian. More precisely, for any rotationally
invariant function $f(|{\hat {y}}|)$ the identity  
\begin{eqnarray}\label{Eq3.56}
  \int_{{\Bbb R}^n} &&
  \frac{ d {\hat{y}}}{(2\pi\Omega)^{n/2}}\:
  \exp\left\{-\frac{1}{2\Omega}({\hat {x}}-{\hat
      {y}})^2\right\}    
  f(|{\hat {y}}|)\nonumber\\
  &&=\left.\exp\left\{\frac{\Omega}{2}
    \left(\frac{d^2}{d\rho^2}
    +\frac{n-1}{\rho}\frac{d}{d\rho}\right)\right\}
  f(\rho)\right|_{\rho=|\hat{x}|} 
\end{eqnarray} 
is valid for all $n\in\Bbb{N}$ 
and its application to (\ref{eq:re9}) yields
\begin{eqnarray}\label{Eq3.58}
  \lefteqn{\varphi^{\Omega}_n(\rho)
  = c\:(2\pi\Omega)^{n/2}}\nonumber\\
  &&\quad\;\;\times\frac{ \displaystyle
    \exp\left\{\frac{\Omega}{2}\, \left(\frac{d^2}{d
          \rho^2}+\frac{n-1}{\rho}\frac{d}{d
          \rho}\right)\right\}
    {\rm e}^{-\rho^2/2 + \varphi^{\Omega}_n(\rho)}}
  {\displaystyle s_n\int_0^{\infty} d \eta\:\eta^{n-1}
    \:{\rm e}^{-\eta^2/2 + \varphi^{\Omega}_n(\eta)}}\,.
\end{eqnarray}
Here $s_n:=n\pi^{n/2}/\Gamma(n/2+1)$ denotes the surface of the
unit sphere in ${\Bbb R}^n$. The normalization (\ref{Eq3.52})
translates to 
\begin{equation}\label{Eq3.59}
  s_n\:\int_0^{\infty} d \rho\:\rho^{n-1}
  \varphi_n^{\Omega}(\rho)=2c\:(2\pi \Omega)^{n/2}\,.
\end{equation}
Observing, that 
$\lim_{n\to 0} \left(n\int_{0}^{\infty} d
  \rho\:\rho^{n-1}f(\rho)\right) = f(0)$ 
is valid for any sufficiently fast decaying function $f$, we get from
(\ref{Eq3.58}) and (\ref{Eq3.55}) 
\begin{equation}\label{Eq3.61}
  \varphi^{\Omega}_0(\rho)=2c\:{\rm e}^{-2c}
  \exp\left\{\frac{\Omega}{2}\left(\frac{d^2}{d
        \rho^2}-\frac{1}{\rho}\frac{d}{d
\rho}\right)\right\}
  {\rm e}^{-\rho^2/2 + \varphi_0^{\Omega}(\rho)}
\end{equation}
and
\begin{equation}\label{Eq3.60}
  G(\Omega)=
  \frac{1}{2\Omega}\:
  {\rm e}^{-2c}\:
  \int_0^{\infty} d \rho\:\rho\:
  {\rm e}^{-\rho^2/2 + \varphi^{\Omega}_0(\rho)}\,,
\end{equation}
respectively. Here we
made use of 
\begin{equation}
  \label{norm}
  \varphi_0^{\Omega}(0)=2c
\end{equation}
 which is obtained from (\ref{Eq3.59}).
The remaining
task is to compute $\varphi_0^{\Omega}(\rho)$ up to first order in
$\Omega$, 
which, in turn, determines the ${\cal O}(1)$-contribution of
$G(\Omega)$.

To this end we make the expansion
\begin{equation}\label{eq:re11}
  \varphi_0^{\Omega}(\rho)=\varphi_0^{0}(\rho)+g(\rho)
  \:\Omega+{\cal O}(\Omega^2)\,,
\end{equation}
insert it into (\ref{Eq3.61}) and equate the zeroth- and first-order
terms
\begin{eqnarray}
&&  \varphi_0^0(\rho)=-W\left(-2c\:{\rm e}^{-2c-\rho^2/2}\right)\,,
   \label{eq:re12} \\
&& \displaystyle g(\rho)= \left[2\left(1-\varphi_0^{0}(\rho)\right)\right]^{-1}
  \left(\frac{d^2}{d
      \rho^2}- \frac{1}{\rho}\frac{d}{d
      \rho}\right)\varphi_0^{0}(\rho) \,.    \label{Eq3.66}
\end{eqnarray}
Here, $W$ denotes the principal branch of the Lambert $W$-function,
\cite{CoGoHa96} which is defined as the real solution of the transcendental
equation 
\begin{equation}
  \label{lambert}
  W(x)\:\exp\bigl(W(x)\bigr)=x\,, \qquad\qquad x> -1/e\,.
\end{equation}
Moreover, from inserting (\ref{eq:re11}) into (\ref{Eq3.60}), we get
the desired expansion
\begin{equation}\label{eq:re13}
  G(\Omega)=\frac{1}{4c}\:\int_0^{\infty} d \rho\:\rho\:
  \varphi_0^0(\rho)\,\left[ \Omega^{-1} + g(\rho)\right] \; + {\cal
O}(\Omega)\,.
\end{equation}
In order to calculate the integrals in (\ref{eq:re13}), we employ the
relation 
\begin{equation}\label{Eq3.63}
  \frac{d\varphi_0^0(\rho)}{d\rho}
  = -\frac{\rho\:\varphi_0^0(\rho)}{1-\varphi_0^0(\rho)}\,,
\end{equation}
which follows from (\ref{eq:re12}) and (\ref{lambert}). It helps to
rewrite both the first term
\begin{equation}\label{Eq3.64}
  \rho\:\varphi_0^0(\rho)=
  \frac{d}{d\rho}
  \:\left[\varphi_0^0(\rho)
    \left(\frac{1}{2}\:\varphi_0^0(\rho)-1\right)\right]
\end{equation}
and, in combination with (\ref{Eq3.66}), the second term  
\begin{eqnarray}\label{eq:re14}
  &&\rho\:\varphi_0^0(\rho)g(\rho)\nonumber\\
  &&=-\frac{1}{2}\frac{d}{d\rho}
  \left[\frac{1}{2}
    \left(\frac{d}{d\rho}
      \varphi_0^0(\rho)\right)^2-\ln\left(1-\varphi_0^0(\rho)\right)-
    \varphi_0^0(\rho)\right] \nonumber\\ 
\end{eqnarray}
of the integrand in (\ref{eq:re13}). Observing in addition the
normalization (\ref{norm}) and $(d\varphi_{0}^{0}/d\rho)(0) =
0$, which also follows from (\ref{Eq3.63}), the evaluation of the
integral in (\ref{eq:re13}) leads to 
\begin{equation}\label{eq:re15}
  G(\Omega)=
  \frac{1-c}{2\Omega}-\frac{1}{8c}\:\big[\ln(1-2c)+2c\big]+{\cal
O}(\Omega) \,.
\end{equation}
Hence, the singular contributions for $\Omega \to 0$ cancel in 
(\ref{eq:re4}), and we end up with the expression
\begin{equation}\label{Eq3.68}
  \lim_{N\to\infty}\langle\eta\rangle=
  \frac{\zeta a^2}{8cd}\left[\ln\!\left(\frac{1}{1-2c}\right)- 2c\right]   
\end{equation}
for the averaged viscosity, which coincides with the exact result
(\ref{Eq3.24}) derived in the previous section.

We conclude that the replica trick and the assumption of replica
symmetry provide an exact method to perform the disorder average in 
the random matrix ensemble (\ref{eq:re2}) below the critical concentration.
This was demonstrated for the viscosity, but we conjecture that it
holds more generally, e.g.\ for the density of states.

\section{The intermediate incoherent scattering function}\label{Sec4}

Dynamic density fluctuations in polymer melts are encoded in the 
intermediate incoherent scattering function (\ref{Eq2.11}). The leading
long-time behaviour of its average $\left\langle
  S_{t}(q)\right\rangle$ over all
crosslink realizations was computed in Ref.\ \onlinecite{BrGoZi97} for
the ensemble (\ref{Eq3.18}) of uncorrelated 
crosslinks, for which $c_{{\rm crit}}=\frac{1}{2}$. Among others,
it was found that 
\begin{itemize}
\item[(1)] 
the incoherent scattering function decays for all crosslink
concentrations $0<c<c_{{\rm crit}}$ like a stretched exponential
\begin{equation} \label{Eq4.01}
\left\langle S_t({q})\right\rangle \stackrel{t\to\infty}{\sim} t^{-1/2} 
\exp\bigl\{-(t/t_q)^{\alpha}\bigr\} \qquad {\rm with~~} \alpha =\frac{1}{2}\,,
\end{equation}
where $t_q \sim q^{-2}$ is a diffusive time scale,
\item[(2)]
as the gelation transition is approached the diffusive time scale diverges
\begin{equation} \label{Eq4.02}
q^2 t_q \stackrel{c\uparrow c_{{\rm crit}}}{\sim} 
(c_{{\rm crit}}-c)^{-\mu} \qquad {\rm with~~} \mu = 2\,, 
\end{equation} 
and the effective diffusion constant vanishes linearly
\begin{eqnarray}  \label{Eq4.03}
D_{{\rm eff}}^{-1} &:=& \lim_{q\to 0} q^2\int_0^{\infty}\! d  t\; 
\left\langle S_t(q)\right\rangle = 
D_{0}^{-1}\sum_{n=0}^{\infty}n^{2}\tau_{n} \nonumber\\
& \stackrel{c\uparrow c_{{\rm crit}}}{\sim} & (c_{{\rm crit}}-c)^{-y}\,, 
\end{eqnarray}
where $y=1$ and $D_{0}:= (\zeta L)^{-1}$ is the bare diffusion constant,
\item[(3)]
the incoherent scattering function decays
algebraically at the critical point 
\begin{equation}  \label{Eq4.04}
\left\langle S_t(q)\right\rangle \stackrel{t\to\infty}{\sim} t^{-x}
\qquad\qquad 
{\rm with~~} x = \frac{1}{2}\,. 
\end{equation}
\end{itemize}
The analysis in Ref.\ \onlinecite{BrGoZi97} relies on the cluster
decomposition (\ref{Eq2.26}) for $S_{t}(q)$ in the form (\ref{Eq2.30})
\begin{equation}
  \label{scluster}
  \left\langle S_{t}(q)\right\rangle = \sum_{n=0}^{\infty} n\tau_{n}
  \left\langle S_{t}(q)\right\rangle_{n} \,.
\end{equation}
In addition, the plausible assumption has been used that the
long-time asymptotics of $\left\langle S_{t}(q)\right\rangle$ is
obtained by inserting the long-time asymptotics of
$\left\langle S_{t}(q)\right\rangle_{n}$ into the right-hand side of
(\ref{scluster}). 

The purpose of this section is twofold. First, for the case of
uncorrelated crosslinks as in Ref.\ \onlinecite{BrGoZi97}, we check
whether the corrections to the leading long-time behaviour of
$\left\langle S_{t}(q)\right\rangle_{n}$ do not influence the leading
long-time behaviour of 
$\left\langle S_{t}(q)\right\rangle$, as was assumed in Ref.\
\onlinecite{BrGoZi97}.  
This is done in Subsection~\ref{Sec4.2} by asymptotic evaluations of
an upper and a lower bound on $\left\langle S_{t}(q)\right\rangle$,
which are 
constructed in Subsection~\ref{Sec4.1}.
Second, in Subsection~\ref{Sec4.3} we determine the exponents $\alpha$,
$\mu$, $y$ and $x$ from the known critical exponents of random bond
percolation by scaling relations. This generalizes the results in Ref.\
\onlinecite{BrGoZi97} to a non-mean-field distribution of crosslinks. 

\subsection{Upper and lower bound on
  $\bbox{\left\langle S_{t}(q)\right\rangle}$}\label{Sec4.1}

Since the connectivity matrix $\Gamma$ is positive
semi-definite, the second exponential in (\ref{Eq2.29}) is bounded
above  by 1, hence
\begin{equation}\label{Eq4.1}
  S_t(q)
  \le
  \sum_{k=1}^{K}\frac{N_k}{N}
  \exp\!\left\{-\frac{q^2t}{\zeta N_kL}\right\}.
\end{equation}
For the computation of the crosslink average it is advantageous to
reorder the sum according to (\ref{Eq2.30})
\begin{equation}\label{Eq4.2}
  \left\langle S_t(q)\right\rangle
  \le {\cal U}_{t}(q) := 
  \sum_{n=1}^{\infty}n\tau_n
  \exp\!\left\{-\frac{D_0q^2t}{n}\right\}\,.
\end{equation}

A lower bound on $S_t(q)$ is obtained by neglecting the double-exponential
contribution in (\ref{Eq2.29})
\widetext
\begin{equation}\label{Eq4.3}
    S_t(q)\ge\sum_{k=1}^{K}
    \frac{N_k}{N}  
    \exp\!\left\{-\frac{q^2t}{\zeta N_kL}\right\}
    \sum_{i\in {\cal N}_k}
    \sum_{s=1}^L\frac{1}{N_kL}
    \exp\!\left\{-\frac{q^2a^2}{d}\frac{1-E_0({\cal N}_{k})}{{\Gamma}({\cal
          N}_{k})}\:(i,s;i,s)\right\}\,. 
\end{equation}
\narrowtext
Next we apply the Jensen inequality to the right-hand side of
(\ref{Eq4.3}) by replacing the normalized $i$- and 
$s$-sums over the exponentials by the exponential of the normalized
sums. Together with (\ref{Eq2.30}) this yields
\begin{eqnarray}\label{Eq4.4}
  \left\langle S_t(q)\right\rangle
  \ge
  \sum_{n=1}^{\infty} &&
  n\tau_n
  \exp\!\left\{-\frac{D_0q^2t}{n}\right\} \nonumber\\
  && \times\;\left\langle \exp\!\left\{-\frac{q^2a^2}{dnL}
      \rm{Tr}\left(\frac{1-E_0}{\Gamma}\right)\right\}\right\rangle_n\,.
\end{eqnarray}
The lower bound is completed by yet another application of the Jensen
inequality, $\left\langle \exp (\bullet)\right\rangle_{n} \ge
\exp(\left\langle \bullet\right\rangle_{n})$, and the identification
(\ref{Eq2.25}) of 
the resulting exponent with the viscosity
\begin{equation}\label{Eq4.4a}
  \left\langle S_t(q)\right\rangle \ge {\cal L}_{t}(q) :=
    \sum_{n=1}^{\infty}  n\tau_n
    \exp\!\left\{- D_0 q^2 \left(\frac{t}{n} + 2 \left\langle
          \eta\right\rangle_{n}\right)\right\} \,. 
\end{equation}
In summary, we have found the chain of inequalities
\begin{equation}
  \label{444.5}
  {\cal L}_{t}(q) \le \left\langle S_{t}(q)\right\rangle \le {\cal U}_{t}(q)\,.
\end{equation}

%
\subsection{Uncorrelated crosslinks}\label{Sec4.2}

It was argued in Ref.\ \onlinecite{BrGoZi97} that $\left\langle
  S_{t}(q)\right\rangle_{n} \stackrel{t\to\infty}{\sim}
\exp\{-D_{0}q^{2}t/n\}$. This relation 
follows from (\ref{Eq2.29}). Assuming that the corrections to this
leading behaviour do not play a r\^{o}le in the cluster decomposition  
(\ref{scluster}), the Kohlrausch law
\begin{eqnarray}
  \label{prlasy}
  \left\langle S_{t}(q)\right\rangle &\stackrel{t\to\infty}{\sim}&
  \sum_{n=0}^{\infty}  
  n\tau_{n}\exp\bigl\{-D_{0}q^{2}t/n\bigr\} \nonumber\\
  &\stackrel{t\to\infty}{\sim}&
  \frac{1}{(8c^2D_0q^2t)^{1/2}}\exp\!\left\{-2[h(c)D_0q^2t]^{1/2}\right\}\,, 
  \nonumber\\   
\end{eqnarray}
$0<c\le\frac{1}{2}$, was found \cite{BrGoZi97} which yields the
critical exponents in 
(\ref{Eq4.01}) -- (\ref{Eq4.04}). Here we introduced the function 
$h(c):= 2c - 1 - \ln(2c)$ and observed $h(c)\sim (c_{{\rm crit}} -
c)^{2}$ for $c\uparrow_{{\rm crit}}$. In order to derive the second line
in (\ref{prlasy}), the explicit form (\ref{Eq3.19}) of $\tau_{n}$ in
the case of uncorrelated crosslinks was used, $n!$ was replaced by its
Stirling asymptotics, and the sum over $n$ by an integral, whose
evaluation yielded the result. 

In this subsection we want to confirm that the corrections to the
asymptotic behaviour of $\left\langle S_{t}(q)\right\rangle_{n}$ do
not influence the 
Kohlrausch law (\ref{prlasy}). To do so we employ the bounds
(\ref{444.5}). We also perform a more careful evaluation of the sums
over the cluster sizes $n$, than the one which was used in Ref.\
\onlinecite{BrGoZi97} to derive the second line in (\ref{prlasy}). Details of
this calculation are deferred to the Appendix. 

We start with the upper bound (\ref{Eq4.2}) which reads in the
notation (\ref{appnot}) of the Appendix as
${\cal U}_{t}(q) = F(0,D_{0}q^{2}t)$. Therefore we conclude from 
(\ref{upper}) that 
\begin{equation}
  \label{upperasy}
  \lim_{q^{2}t\to\infty} \frac{{\cal U}_{t}(q)}{(8c^{2}D_{0}
    q^{2}t)^{-1/2}\exp\bigl\{-2[h(c)D_{0}q^{2}t]^{1/2}\bigr\}} =1 \,.
\end{equation}
This result also confirms the previously obtained asymptotic equality
in the second line of (\ref{prlasy}). 

As to the lower bound (\ref{Eq4.4a}) we first recall the expression
(\ref{Eq3.29}) for $\left\langle \eta\right\rangle_n$ and the inequality 
$\left\langle \eta\right\rangle_{n}\big|_{L=1} \le b \sqrt{n}$, which
follows from 
(\ref{etan1}) and (\ref{Eq3.21}), with $b>0$ being a finite constant. 
This yields, again in the notation (\ref{appnot}) of
the Appendix, 
\begin{equation}
  {\cal L}_{t}(q) \ge {e}^{-B_{1}q^{2}}\, F(q^{2}B_{2}, D_{0}q^{2}t)\,,
\end{equation}
where $B_{1} := D_{0}\zeta l^{2}L{\cal Q}_{L}/(2d)$ and 
$B_{2} := D_{0}b\zeta a^{2}L(1+{\cal Q}_{L}l^{2}/a^{2})/(2d)$.
Consequently, (\ref{lower}) implies 
\begin{equation}
  \label{lowerasy}
  \lim_{q^{2}t\to\infty \atop q^{6}t\to 0}
  \frac{{\cal L}_{t}(q)}{(8c^{2}D_{0}
    q^{2}t)^{-1/2}\exp\bigl\{-2[h(c)D_{0}q^{2}t]^{1/2}\bigr\}} \ge 1 \,.
\end{equation}
Taken together, (\ref{upperasy}) and (\ref{lowerasy}) provide us with
the equality 
\begin{equation}
  \lim_{q^{2}t\to\infty \atop q^{6}t\to 0}
  \frac{\left\langle S_{t}(q)\right\rangle}{(8c^{2}D_{0}
    q^{2}t)^{-1/2}\exp\bigl\{-2[h(c)D_{0}q^{2}t]^{1/2}\bigr\}} = 1 \,.
\end{equation}
In other words, the bounds (\ref{444.5}) only confirm the previously
obtained Kohlrausch law (\ref{prlasy}), if one also considers the
limit of small momentum transfer--- as may be a reasonable
approximation for an experimental low-angle-scattering situation. 
Nevertheless, we believe that the stretched exponential (\ref{prlasy})
reflects the true long-time asymptotics of $\left\langle
  S_{t}(q)\right\rangle$  
for finite $q$, too, and that the lower bound (\ref{Eq4.4a}) is not sharp
enough to reproduce this.  

On the other hand the bounds (\ref{444.5}) are sharp enough to
establish the expression 
\begin{equation}
  \label{deeff2}
  D_{{\rm eff}}^{-1} = D_{0}^{-1}\sum_{n=0}^{\infty}n^{2}\tau_{n}
  = \frac{1}{D_{0}(1-2c)} 
\end{equation}
for the effective diffusion constant $D_{{\rm eff}}$. The first equality was
already stated in (\ref{Eq4.03}). The second follows for
$0<c<\frac{1}{2}$ from differentiating the relation
$1=\sum_{n=1}^{\infty} n\tau_{n}$ with respect to the crosslink
concentration $c$. Obviously, the result is in accordance with $y=1$ in
(\ref{Eq4.03}). The linear dependence of $D_{{\rm eff}}$ on $c$ is
displayed in Figure~\ref{fig1} in the Introduction.

%
\subsection{Scaling description}\label{Sec4.3}

In this subsection we extend the results of Ref.\
\onlinecite{BrGoZi97} to the 
crosslink statistics generated by $d$-dimensional random bond
percolation, whose basic properties were recalled in Subsection \ref{Sec3.2}. 

For the case of uncorrelated crosslinks we have seen in Ref.\
\onlinecite{BrGoZi97}---and confirmed in the preceeding subsection by
alternative methods---that
the critical exponents (\ref{Eq4.01}) -- (\ref{Eq4.04}) of the
Kohlrausch law (\ref{prlasy}) do not depend on the number $L$ of
monomers per polymer. Therefore we restrict ourselves to the case
$L=1$ of Brownian particles in the following. The results of the last
subsection also motivate us to replace $\left\langle
  S_{t}(q)\right\rangle_{n}$ by its 
long-time asymptotics in (\ref{scluster}), that is,
\begin{equation}
  \label{scas}
  \left\langle S_{t}(q)\right\rangle \stackrel{t\to\infty}{\sim}
  \sum_{n=0}^{\infty}  
  n\tau_{n}\exp\bigl\{-D_{0}q^{2}t/n\bigl\}\,.
\end{equation}
For $t\to\infty$ this series is expected to be dominated by terms with
large $n$. Therefore we insert the scaling law (\ref{Eq3.7}) for
$\tau_{n}$ in (\ref{scas}) and use the fact that the associated
scaling function $f$ decays exponentially for large $n$. This yields
up to a constant
\begin{equation} \label{sumreplace}
  \left\langle S_{t}(q)\right\rangle \stackrel{t\to\infty}{\sim}
  \sum_{n=0}^{\infty} n^{1 -\tau} \exp\bigl\{ -(c_{{\rm crit}} -
  c)^{1/\sigma} n - D_{0}q^{2}t/n\bigr\}\,.
\end{equation}
In order to proceed, we replace the sum over $n$ in (\ref{sumreplace})
by the integral
\begin{equation}
  \label{prelaplace}
  \int_{0}^{\infty}\! d  n\; n^{1-\tau} \exp\bigl\{ -(c_{{\rm crit}} -
  c)^{1/\sigma} n - D_{0}q^{2}t/n\bigr\}\,,
\end{equation}
which should not affect the long-time asymptotics either. Using
Formula 3.471.9 in Ref.\ \onlinecite{GrRh80}, we then express the integral in
(\ref{prelaplace}) in terms of a Bessel function. Its asymptotic
expansion according to Formula 8.451.6 in Ref.\ \onlinecite{GrRh80}
finally gives, up to a constant,
\begin{eqnarray}
  \label{scar}
  \lefteqn{
  \left\langle S_{t}(q)\right\rangle \stackrel{t\to\infty}{\sim} 
  \frac{(c_{{\rm crit}} -  c)^{(2\tau -5)/(4\sigma)}}%
  {(D_{0}q^{2}t)^{(2\tau -3)/4}} }\nonumber\\  
  &&\hspace*{2cm}\times\;\exp\bigl\{ -2[D_{0}q^{2} (c_{{\rm crit}}
  -  c)^{1/\sigma}t]^{1/2}\bigr\}
\end{eqnarray}
\narrowtext
for $c<c_{{\rm crit}}$. For $c=c_{{\rm crit}}$ we conclude 
\begin{equation}
  \left\langle S_{t}(q)\right\rangle \stackrel{t\to\infty}{\sim}
  \frac{{\rm const.}} 
  {(D_{0}q^{2}t)^{\tau -2}}
\end{equation}
directly from (\ref{prelaplace}).
Hence, we find $\alpha =1/2$ for the critical exponent (\ref{Eq4.01}),
the same value as was found for uncorrelated crosslinks. In contrast,
the exponents defined by the relations (\ref{Eq4.02}) --
(\ref{Eq4.04}) turn out to be different. More precisely, we get the
scaling relations
\begin{eqnarray}
  \mu &=& 1/\sigma         \approx  2.22\,, \\
  y   &=& (3-\tau)/\sigma  \approx  1.82\,, \\
  x   &=& \tau -2          \approx  0.18\,. 
\end{eqnarray}
The approximate numerical values are those for random bond percolation
in three dimensions.\cite{StAh94} A discussion of these results in comparison
to the experimental data will be given in the next section.

\section{Concluding remarks}\label{Sec5}

Starting from a Rouse model of a crosslinked polymer melt, we
discuss the critical dynamics of the gelation transition with
particular emphasis on the static shear viscosity and the long-time
behaviour of the incoherent scattering function. Our main result is an
exact relation for the critical exponent of the viscosity, 
\begin{equation}
  \label{resscal}
  k=\phi-\beta\,,
\end{equation}
in terms of the crossover and thermal exponents of percolation
theory. 
Two crosslink distributions have been analyzed in detail:
one corresponding to a mean-field percolation model and one
corresponding to finite-dimensional percolation. For the first
distribution we were able to derive the exact expression
(\ref{Eq3.30}) for the static shear viscosity. This result is valid
for all crosslink concentrations $0<c<c_{{\rm crit}}=\frac{1}{2}$
and exhibits a logarithmic divergence at the critical crosslink
concentration $c_{{\rm crit}}$, that is, $k=0$. The critical
exponent $k=0$ also follows from the scaling relation (\ref{resscal})
when inserting the mean-field values $\phi=1$, see e.g.\ Ref.\
\onlinecite{StJaOe99}, and  
$\beta =1$. For the second crosslink distribution the scaling relation
(\ref{resscal}) can either be evaluated in terms of the 
$\epsilon=6-d$ expansion \cite{StJaOe99} for $\phi$ and $\beta$,
yielding $k=\epsilon/6+11\epsilon^2/1764+{\cal
  O}(\epsilon^3)$. Alternatively, one may insert the numerical value
of $\phi$ known from high-precision simulations of random
resistor networks in $d=3$ dimensions, \cite{GiLo90} which, together
with the corresponding value for $\beta$ gives
$k|_{d=3}\approx 0.71$.    

The experimental values for $k$ are systematically larger. Early
experiments by Adam {\it et al.\/} \cite{AdDeDuHiMu81-AdDeDu85} were
performed near the gelation threshold of polycondensation reactions.
For the samples with low molecular weight, data for
the viscosity were fitted to a power-law divergence with exponent
$k=0.8 \pm 0.1$. On the other hand, the viscosity data of samples with
high molecular weight could not be fitted to a power-law
divergence. Considerably larger exponent values were observed in more
recent experiments on silica gels and epoxy raisins. Colby {\it et al.\/}
\cite{CoGiRu93} report values $1.4\leq k \leq 1.7$ and Martin {\it et al.\/}
obtained $k=1.4\pm 0.2$ from viscoelastic measurements \cite{MaAdWi88} and
$k=1.5\pm 0.2$ rather indirectly from measurements of the incoherent
scattering function. \cite{MaWi88-AdMa90-MaWiOd91} The origin of the
scatter of the 
experimental data is not clear. A splitting of the static universality
class into different dynamic ones has been suggested.
\cite{AdDeDuHiMu81-AdDeDu85}
Another possible explanation is the size of the critical region, which
is expected to depend on chain length. \cite{Ge77}
The observed range of exponent values could then be interpreted as a
crossover from mean-field to critical behaviour.

What are the shortcomings of our theory?  We have written the average
macroscopic viscosity as a weighted sum of contributions from
individual clusters. Such a decomposition holds exactly within the
Rouse model of a crosslinked melt, but is expected to be valid more
generally---as long as there are no interactions between different
clusters. Retaining the cluster decomposition, the only unknown is the
contribution of an individual cluster, because the distribution of
clusters should indeed be given by percolation theory and hence is
known to high precision. As far as single clusters are concerned, we
expect that relaxation times are longest for Rouse dynamics. If, for
example, 
we were to use Zimm dynamics together with a pre-averaging
approximation, we would find shorter relaxation times as compared to
the Rouse model and hence even smaller values of $k$. This suggests
that the discrepancy between theory and experiment cannot be cured, if
we retain the cluster decompostion. Hydrodynamic interactions,
excluded-volume interactions and entanglement effects, all invalidate
the cluster decomposition. One expects entanglement effects to play a
vital r\^ole in stress relaxation. \cite{DoEd85} However, the static
viscosity measures stress relaxation only on the longest time scales
in the sol phase. It has been argued
\cite{CoGiRu93,RuZuMcBa90-MiMcYoHaJo98} that in the regime close to
the transition, entanglement effects are less important because of two
reasons: first, there are almost no permanent entanglements in the
sense of interlocking loops. Second, the time scale of a temporary
entanglement of two clusters is determined by the smaller cluster,
whereas the dynamics on the longest time scales is determined by the
larger cluster. From calculations of static quantities we know that
the excluded-volume interaction is essential for the stability of the
network in the gel phase. \cite{GolCasZip96}  In the sol phase, on the other
hand, the r\^ole of the excluded-volume interaction is less transparent.
On one hand, the $\epsilon =6-d$ expansion for a gelation model with
excluded-volume interactions \cite{GoPe00} yields the same critical
exponents as obtained from the $\epsilon$-expansion of percolation
theory.\cite{StAh94} 
On the other hand, the excluded-volume
interaction is known to be relevant \cite{Sch99} for static and dynamic
properties of polymer melts and solutions below $d=4$.

The decay of density fluctuations in silica gels has been investigated
by quasi elastic light scattering. \cite{MaWi88-AdMa90-MaWiOd91}  In
the sol phase, a stretched exponential of the autocorrelation was
observed, $S_t(q) \sim \exp\{ -(t/t_q)^{\alpha}\}$ with $\alpha=0.65
\pm 0.05$.  The experimentally determined timescale $t_q$ is diffusive
and diverges as the gelation transition is approached $t_q\sim
(c_{{\rm crit}}-c)^{-2.2}$. The critical behaviour of the diffusion
constant was determined as well and in particular the exponent value
$y=1.9 \pm 0.1$ was found. At the critical point, the scattering
experiments reveal an algebraic decay in time of $S_t(q)$ with an
exponent $x=0.135 \pm 0.015$. All these findings are in qualitative
agreement with our theory. In fact our expression (\ref{Eq2.29}) for
$S_t(q)$ has been suggested on phenomenological grounds as a starting
point for the discussion of the critical dynamics at the sol-gel
transition. \cite{Ge79} The exponent values of mean-field theory, see
Eqs.\ (\ref{Eq4.01}) -- (\ref{Eq4.04}),
deviate from the experimental ones, as one would expect. If we use the
scaling description of finite-dimensional percolation, then the
exponent $\alpha$ of the stretched exponential is unchanged. Its
value, $\alpha=1/2$, is characteristic of Rouse dynamics and
independent of cluster statistics. The corrections to the other
exponents go in the right direction, see Eqs.\ (\ref{Eq4.04}). However,
discrepancies beyond the experimental error bars remain and are
probably due to our neglect of excluded-volume and hydrodynamic
interactions.

Our approach can be extended to study stress relaxation at finite
frequencies. Within the Rouse model, the dynamics is completely
determined by the eigenvalue spectrum of the random
connectivity matrix $\Gamma$. For the mean-field distribution of
crosslinks the eigenvalue spectrum can be computed with the replica
trick \cite{RoBr88} so that stress relaxation at finite
frequencies becomes accessible to analytical theory. Work along these
lines is in progress.

\begin{acknowledgments}
This work was supported by the DFG through SFB 345. K.B.\ acknowledges
financial support by the DFG under grant No.\ Br 1894/1-1.
\end{acknowledgments}

\appendix
\section*{Evaluation of the bounds (\protect\ref{444.5})} 

Here we present some technical details which are needed in
the asymptotic evaluation for $t\to\infty$ of the bounds (\ref{444.5})
on the averaged intermediate incoherent scattering function 
$S_{t}(q)$. 

The quantity of interest is the series
\begin{equation}
  \label{appnot}
  F(B,T) := \sum_{n=1}^{\infty} n \tau_{n}\, {e}^{-T/n} {e}^{-B
    \sqrt{n}}\,, 
\end{equation}
where $B\ge 0$, $T>0$ and $\tau_{n} = n^{n-2}{e}^{-nh(c)}/(2c\, n!\,
 {e}^{n})$ follows from (\ref{Eq3.19}) with 
$h(c):= 2c -1-\ln(2c)$. Now, pick a natural number $P$, split the
series in two parts given by the first $P-1$ terms and the terms with 
$n\ge P$, respectively, and apply the Stirling approximation, 
Eq.\ (6.1.58) in Ref.\ \onlinecite{AbSt72},
\begin{equation}
  n! = (2\pi)^{1/2}\, n^{n+1/2}\, \exp\bigl\{-n +\theta(n)/(12n)\bigr\}\,,
\end{equation}
$0 < \theta(n) < 1$, to the terms with $n\ge P$. Thus, we infer the
existence of a constant  
\begin{equation}
  \label{upebound}
  {e}^{-1/(12P)} \le u_{P} \le 1
\end{equation}
such that 
\begin{equation}
  \label{upe}
  F(B,T) = R_{P}(B,T) + \frac{u_{P}}{2c \,(2\pi)^{1/2}} \,\tilde{F}(B,T)\,,
\end{equation}
where 
\begin{eqnarray}
  R_{P}(B,T) &:=& \sum_{n=1}^{P-1} \left( n\tau_{n} - \frac{u_{P}}{2c \,
      (2\pi)^{1/2}\, n^{3/2}} \, {e}^{-nh(c)}\right) \nonumber\\
  && \qquad\times\, {e}^{-T/n}\,{e}^{-B\sqrt{n}}
\end{eqnarray}
and
\begin{equation}
  \tilde{F}(B,T) := \sum_{n=1}^{\infty} n^{-3/2} \exp\bigl\{ -nh(c) - T/n -
  B\sqrt{n}\bigr\}\,.
\end{equation}
Note that 
\begin{equation}
  \label{errub}
  |R_{P}(B,T)| \le {e}^{-T/(P-1)} \sum_{n=1}^{P-1}\bigg| n\tau_{n} 
    - \,\frac{u_{P}\,{e}^{-nh(c)}}{2c \,(2\pi)^{1/2}\, n^{3/2}} \bigg|
\end{equation}
decays exponentially for $T\to\infty$, uniformly in $B\ge 0$.

In order to proceed with $\tilde{F}(B,T)$ we use the Fourier
representation 
\begin{equation}
  {e}^{-\alpha/2} = \int_{\Bbb{R}}\! d  x\; {e}^{ i\, x}\;
  \frac{{e}^{-x^{2}/(2\alpha)}}{(2\pi\alpha)^{1/2}} \,, \qquad\quad
  \alpha >0\,, 
\end{equation}
and arrive at 
\begin{equation}
  \label{hansen}
  \tilde{F}(B,T) = \int_{\Bbb{R}}\frac{ d  x}{(4\pi D_{0}t)^{1/2}}\;
  {e}^{ i\, x}\int_{A(x)}^{\infty}\! d \lambda \sum_{n=1}^{\infty} 
  {e}^{-\lambda n}\,{e}^{-B\sqrt{n}}\,,
\end{equation}
where $A(x):=h(c)+x^{2}/(4T)$. According to Eq.\ (11.1.1) in Ref.\
\onlinecite{Han75} the series in (\ref{hansen}) admits an
integral representation such that 
\begin{eqnarray}
  \label{posthansen}
  \tilde{F}(B,T) &=& \frac{B}{4\pi\, T^{1/2}} \int_{\Bbb{R}}\! d  x\;
  {e}^{ i\, x}\int_{A(x)}^{\infty}\! d \lambda \nonumber\\
  && \qquad\qquad \times
  \int_{0}^{\infty}\! d \xi\; \frac{{e}^{-B^{2}/(4\xi)}}{\xi^{3/2}
    ({e}^{\xi +\lambda} -1)}\,.
\end{eqnarray}
After a partial integration with respect to $x$ and the subsequent
changes-of-variables $z:=x(4T)^{-1/2}$ and $\zeta := (2\xi)^{-1/2}$, 
Eq.\ (\ref{posthansen}) reads
\begin{equation}
  \tilde{F}(B,T) = \frac{- i\,}{(\pi T)^{1/2}} \int_{\Bbb{R}}\! d \zeta\;
  \frac{{e}^{-\zeta^{2}/2}}{(2\pi)^{1/2}}\int_{\Bbb{R}}\! d  z\;
  \frac{z\,{e}^{2 i\, z\sqrt{T}}}{{e}^{z^{2}+h(c,B/\zeta)} -1}
\end{equation}
with $h(c,r):= h(c) + r^{2}/2$. The $z$-integration can be done with
the help of the residue theorem. To this end we close the integration
contour with a semicircle in the complex upper half-plane and remark
that the integrand has simple poles $z_{n}(B/\zeta):=
z_{n}'(B/\zeta) +  i\, z_{n}''(B/\zeta)$, $n\in{\Bbb Z}$, in this
half-plane whose real and imaginary parts are given by
\begin{eqnarray}
  z_{n}'(r) &:=& \frac{{\rm sgn}(n)}{\sqrt{2}}\,
  \left[\sqrt{\bigl(h(c,r)\bigr)^{2} + (2\pi 
      n)^{2}} - h(c,r)\right]^{1/2}\,,\nonumber\\
  z_{n}''(r) &:=& \frac{1}{\sqrt{2}}\,
  \left[\sqrt{\bigl(h(c,r)\bigr)^{2} + (2\pi 
      n)^{2}} + h(c,r)\right]^{1/2}\,. \nonumber\\
\end{eqnarray}
Hence, we find 
\begin{eqnarray}
  \label{efftil}
  \tilde{F}(B,T) &=& \left(\frac{\pi}{T}\right)^{1/2} \int_{\Bbb{R}}\! d \zeta\;
  \frac{{e}^{-\zeta^{2}/2}}{(2\pi)^{1/2}} \; \bigg( {e}^{-2[T
      h(c,B/\zeta)]^{1/2}} \nonumber\\
    && \hspace*{2.7cm}+ 2 \tilde{F}_{1}(B/\zeta, T)\bigg)\,,
\end{eqnarray}
where
\begin{equation}
  \tilde{F}_{1}(r,T) := \sum_{n=1}^{\infty} {e}^{-2
    z_{n}''(r)\sqrt{T}} \,\cos[2 z_{n}'(r)\sqrt{T}]\,.
\end{equation}
Using $z_{n}''(r)\ge z_{n}''(0)$, we get the $r$-independent upper
bound
\begin{eqnarray}
  \label{eff1ub}
  |\tilde{F}_{1}(r, T)| &\le&  {e}^{-2 z_{1}''(0)\sqrt{T}}
  \sum_{n=1}^{\infty} {e}^{-2[z_{n}''(0) - z_{1}''(0)]\sqrt{T}}
  \nonumber\\
  & \le & f\, {e}^{-2z_{1}''(0)\sqrt{T}}\,.
\end{eqnarray}
Here, $f$ is a uniform constant for all $T\ge T_{0}>0$, because
$z_{n}''(0) > z_{1}''(0)$ for $n > 1$ and $z_{n}''(0)\sim n^{1/2}$ for
$n\to\infty$. Thus we conclude from (\ref{upe}), (\ref{errub}),
(\ref{efftil}) and (\ref{eff1ub}) that
\begin{equation}
  \label{preupper}
  \lim_{T\to\infty} \frac{F(0,T)}{(8c^{2}T)^{-1/2}\exp\bigl\{-2
    [h(c)T]^{1/2}\bigr\}}   = u_{P}
\end{equation}
holds for all $P\in{\Bbb N}$.

For $B>0$ the inequality $\sqrt{h(c,r)} \le \sqrt{h(c)}
+ 2^{-1/2}|r|$ leads to the lower bound
\begin{eqnarray}
\lefteqn{
\int_{\Bbb{R}}\! d \zeta\; \frac{{e}^{-\zeta^{2}/2}}{(2\pi)^{1/2}}\;
{e}^{-2 [T h(c,B/\zeta)]^{1/2}}} \nonumber\\  
&& \hspace*{.6cm} \ge {e}^{-2[T h(c)]^{1/2}}  
\int_{\Bbb{R}}\! d \zeta\; \frac{{e}^{-\zeta^{2}/2}}{(2\pi)^{1/2}}\;
{e}^{-(2B^{2}T)^{1/2}/|\zeta|}\,.\;\;
\end{eqnarray}
This bound, together with the same arguments which led to
(\ref{preupper}), now yield
\begin{equation}
  \label{prelower}
  \lim_{T\to\infty \atop B^{2}T\to 0} 
  \frac{F(B,T)}{(8c^{2}T)^{-1/2}\exp\bigl\{-2 [h(c)T]^{1/2}\bigr\}}
  \ge u_{P}
\end{equation}
for all $P\in{\Bbb N}$. Letting $P\to\infty$ and observing
(\ref{upebound}), Eq.\ (\ref{preupper}), respectively
(\ref{prelower}), implies
\begin{equation}
  \label{upper}
  \lim_{T\to\infty} \frac{F(0,T)}{(8c^{2}T)^{-1/2}\exp\bigl\{-2
    [h(c)T]^{1/2}\bigr\}}   = 1\,,
\end{equation}
respectively
\begin{equation}
  \label{lower}
  \lim_{T\to\infty \atop B^{2}T\to 0} 
  \frac{F(B,T)}{(8c^{2}T)^{-1/2}\exp\bigl\{-2 [h(c)T]^{1/2}\bigr\}}
  \ge 1\,.
\end{equation}

\widetext
\end{document}